\documentclass[twocolumn]{IEEEtran2}

\usepackage{amscd}
\usepackage{epsfig}

\begin{document}

\newcommand\beq{\begin{equation}}
\newcommand\eeq{\end{equation}}
\newcommand\bea{\begin{eqnarray}}
\newcommand\eea{\end{eqnarray}}
\def\be{\begin{equation}}
\def\ee{\end{equation}}

\def\<{\langle}
\def\>{\rangle}
\def\lbL{ \left[\rule{0pt}{2.4ex} }
\def\rbL{ \right] }
\def\lbm{ \left[\rule{0pt}{2.1ex}\right. }
\def\rbm{ \left.\rule{0pt}{2.1ex}\right] }
\def\eps{\epsilon}
\newcommand{\sig}[2]{\sigma_{#1}^{({#2})}}
\newcommand{\ket}[1]{| #1 \rangle}
\newcommand{\bra}[1]{\langle #1 |}
\newcommand{\braket}[2]{\langle #1 | #2 \rangle}
\newcommand{\proj}[1]{| #1\rangle\!\langle #1 |}
\newcommand{\ba}{\begin{array}}
\newcommand{\ea}{\end{array}}
\newtheorem{theo}{Theorem}
\newtheorem{defi}{Definition}
\newtheorem{lem}{Lemma}
\newtheorem{exam}{Example}
\newtheorem{prop}{Property}
\newtheorem{propo}{Proposition}
\newtheorem{cor}{Corollary}
\newtheorem{conj}{Conjecture}

\setcounter{page}{1}

\title{Quantum Data Hiding}
\author{David P. DiVincenzo, Debbie W. Leung and Barbara M. Terhal
\thanks{IBM Watson Research Center, P.O. Box 218, Yorktown Heights, NY
10598, USA}}

\markboth{Submitted to IEEE Transactions On Information Theory, March 2001}
{Data Hiding}

\date{\today}

\maketitle

\begin{abstract}
We expand on our work on {\em Quantum Data Hiding}~\cite{tdl:prl} --
hiding classical data among parties who are restricted to performing
only local quantum operations and classical communication (LOCC).  We
review our scheme that hides one bit between two parties using Bell
states, and we derive upper and lower bounds on the secrecy of the
hiding scheme.  We provide an explicit bound showing that multiple
bits can be hidden bitwise with our scheme.  We give a preparation of
the hiding states as an efficient quantum computation that uses at
most one ebit of entanglement.  A candidate data hiding scheme that
does not use entanglement is presented.  We show how our scheme for
quantum data hiding can be used in a conditionally secure quantum bit
commitment scheme.
\end{abstract}


\section{Introduction}

It is well known that composite quantum systems can exhibit a variety
of nonlocal properties. When two systems are entangled, as when two
spins are described by the singlet state
$\frac{1}{\sqrt{2}}(\ket{01}-\ket{10})$, local measurements on the two
particles separately can exhibit statistics unexplainable by local
hidden variable theories, such as a violation of Bell's
inequalities~\cite{bell}. An information-theoretic or computational
expression of this feature is that entangled states can function as
nontrivial resources in quantum communication
protocols~\cite{cleve&buhrman:subs}, for example reducing the
amount of classical communication needed to perform certain
distributed computations.

It has been found that even states without quantum entanglement can
exhibit properties of nonlocality that are not present in purely
classical systems.  The first explorations in this direction were
carried out by Peres and Wootters~\cite{peres&wootters}, who studied
the measurements that could optimally distinguish three nonorthogonal
quantum states of which two parties, Alice and Bob, both possess a
single copy. They found that any measurement that can be performed
using a sequence of local operations supplemented by classical
communication between the parties (denoted as LOCC) is not able to
retrieve as much information as a global measurement carried out on
the joint system. Thus, even though no entanglement is present in this
system, the states exhibit nonlocality with respect to their
distinguishability. In Ref.~\cite{qne}, where the term `quantum
nonlocality without entanglement' was coined, a similar phenomenon was
exhibited: it is impossible to use LOCC to perfectly distinguish nine
orthogonal bipartite product states, which are perfectly
distinguishable when nonlocal actions are allowed.

The results that we have presented in Ref.~\cite{tdl:prl}, and on
which we expand in the present paper, can be viewed as the strongest
possible separation between the power of LOCC versus global operations
for the task of distinguishing quantum states.  We use the nonlocality
of our quantum states to establish a protocol of quantum data hiding:
a piece of classical data is hidden from two parties, who each share a
part of the data and are allowed to communicate classically.  Such a
scheme is nontrivial in several respects. First, it is impossible in a
purely classical world.  Second, it is impossible if the state shared
by the two parties is a {\em pure} quantum state. This observation
follows from the result by Walgate {\em et
al.}~\cite{walgate:orthodist} which shows that any two orthogonal
bipartite pure quantum states can be perfectly distinguished by LOCC.
Third, the scheme is extremely secure; it is possible to make the
amount of information obtainable by the parties arbitrarily small; the
number of qubits needed is only logarithmic in the information bound.

The quantum data hiding scheme is secure if the parties Alice and Bob
cannot communicate quantum states and do not share prior quantum
entanglement. In what kind of situations can these conditions be met,
and is our scheme of interest?  One can imagine a situation in which a
third party (the boss) has a piece of data on which she would like
Alice and Bob (some employees) to act by LOCC without the sensitive
data being revealed to them. We have to assume that the boss controls
(1) the channel which connects the two parties and (2) the labs in
which the employees operate, so that the boss can use dephasing to
prevent the quantum communication and to sweep those labs clean of any
entanglement prior to operation.  Our scheme is such that at some
later stage, the boss can provide the employees with entanglement to
enable them to determine the secret with certainty.  This feature is
used for a conditionally secure bit commitment scheme.

An additional advantage of our scheme, besides its
information-theoretic security, is that it can be implemented
efficiently; the number of computation steps required, both classical
and quantum, grows no faster than a polynomial of the input size.  We
find an efficient algorithm to prepare the data hiding states which
also minimizes the use of quantum entanglement.  The algorithm hinges
on a surprising connection between an operation known as the Full
Twirl~\cite{bdsw} and a Twirl over the Clifford
group~\cite{gotthesis}.  The Full Twirl is an important operation in
the study of entanglement while the Clifford group is an important
discrete group in the theory of quantum error correction.

An original goal in our investigations was to establish a data hiding
scheme in which a bit could be hidden from LOCC observers, {\em and}
the data hiding states are unentangled; this would have formed an
extremely strong example of `nonlocality without entanglement'.
Separable (unentangled) hiding states are interesting also because the
security for hiding a single bit implies directly that hiding
independently distributed multiple bits is also secure.
The quantum data hiding protocol using Bell states does not entirely
achieve this goal, since it still requires a small amount of
entanglement. 
We propose an alternative quantum data hiding scheme that uses
unentangled hiding states. We can only rigorously analyze this scheme
for small systems, but on the basis of this analysis, we conjecture
this scheme is secure, in the same way as the scheme using Bell
states.

Our paper is organized in the following way. In Section \ref{sec:povm}
we review the general setup that is needed to analyze the problem of
distinguishing a pair of states by LOCC. We derive a condition
for any LOCC measurement that attempts to distinguish a pair of
states. A related condition has also been discussed in
Ref.~\cite{entop:cirac}. 
In Sections~\ref{sec:bithid1}-\ref{sec:multiple} we discuss
various aspects of the security of our quantum data hiding protocol:
In Section~\ref{sec:bithid1} we discuss the scheme for hiding a single
bit. This scheme was first presented and proved secure in
Ref.~\cite{tdl:prl}. Our analysis here goes into more detail.
Furthermore, in Section~\ref{tightb} we show that the bound on the
retrievable information is fairly tight -- we find a simple LOCC
measurement that retrieves an amount of information close to our
proved bound.
In Section~\ref{sec:genlower}, we digress to show a general result, on
how well a single bit can be hidden in two {\em arbitrary} orthogonal
bipartite states.  We obtain a lower bound on the retrievable information, 
which shows in another way that our scheme has nearly optimal hiding
capability.
In Section~\ref{sec:multiple} we extend our protocol to hide $k >1$
bits.  We are able to prove a good upper bound on the information
retrievable by LOCC, exploiting the symmetry of our hiding states.  

In Sections~\ref{sec:prepstate}-\ref{sec:disc}, we present various 
schemes and discussions related to our quantum data hiding protocol: 
In Section~\ref{sec:prepstate} we present an efficient algorithm to
prepare the data hiding states which also minimizes the use of quantum
entanglement.  We also prove the equivalence of the Full Twirl and 
the Twirl over the Clifford group.  
In Section~\ref{sec:altsep}, we discuss the reason that the security
for hiding a single bit implies the security for hiding independently
distributed multiple bits, and we describe an alternative scheme for
hiding bits that uses unentangled hiding states.
In Section~\ref{sec:qbc}, we apply the quantum data hiding scheme to
construct a conditionally secure quantum bit commitment protocol. We
conclude our paper with some discussion and open questions in
Section~\ref{sec:disc}.  

The discussion up to Section~\ref{sec:secpf1} is a prerequisite for
all other Sections, which can then be read independently.
Throughout the paper, the tensor product of two $d$-dimensional
Hilbert spaces is denoted as ${\cal H}_d \otimes {\cal H}_d$, and a
positive semidefinite matrix or operator $A$ (with nonnegative
eigenvalues) is denoted as $A \geq 0$.

\section{General formalism for operations to learn the secret}
\label{sec:povm}

In quantum mechanics, a large class of state changes can be described 
using the formalism of quantum operations.  
A quantum operation is a completely positive
map~\cite{Choi75a,Nielsen00} on operators in a Hilbert space $\cal H$.
A convenient representation of a quantum operation is the {\em
operator-sum representation}~\cite{Schumacher96a,Choi75a}:
\beq
	{\cal S}[\rho] = \sum_k S_k \rho S_k^\dagger\,,
\label{osr}
\eeq
where $S_k$ are operators acting on $\cal H$.  
We restrict our discussion to {\em trace preserving} quantum
operations, for which $\sum_k S_k^\dagger S_k = I$.  The adjoint of
the quantum operation $\cal S$ is ${\cal S}^\dagger$, whose action can
be expressed as ${\cal S}^\dagger[\rho]=\sum_kS_k^\dagger\rho S_k$.  A
state is represented by a density operator $\rho \geq 0$ with unit
trace.  A rank one density matrix, $\rho = |\psi\>\<\psi|$, is called
{\em pure} and is often represented as a Hilbert-space vector
$|\psi\>$.

We will consider bipartite density matrices held by two parties Alice
and Bob.  Peres and Horodecki {\em et al.}~\cite{Asher96,nec_horo}
have introduced a test for the separability of such bipartite density
operators, which we will use throughout this paper.  Their criterion
is satisfied by a density matrix $\rho$ when $({\bf 1}_A \otimes
T_B)[\rho]\geq 0$, where $T_B$ stands for matrix transposition in any
chosen basis for Bob's Hilbert space, and ${\bf 1}_A$ is the identity
operation on Alice's Hilbert space.  We will say that such a density
matrix $\rho$ is PPT, {\em positive} under {\em partial
transposition}.

Our goal is to hide classical data in bipartite mixed states,
meaning that Alice and Bob cannot learn the
secret if they do not share entanglement and can only perform quantum
operations in the LOCC class.
An LOCC quantum operation $\cal S$ has the Peres-Horodecki property,
or {\em P-H property}\footnote{Rains calls operations with this
property {\em p.p.t. superoperators}; see Section~\ref{sec:disc} for
further discussion.}: if $\rho$ is PPT then $({\bf
1}_{A2,B2}\otimes{\cal S}_{A1,B1})[\rho]$ is also PPT.  The subscripts
in this expression emphasize that $\cal S$ may act only on part of the
bipartite Hilbert space $A$1,$A$2/$B$1,$B$2 on which $\rho$ exists.
(This extension to larger Hilbert space parallels the definition of
complete positivity of quantum operations.)
While the LOCC class is highly non-trivial to
characterize~\cite{rainssep}, the P-H property itself is much simpler
to check, and we will use this to derive necessary conditions for LOCC
operations.
In our analysis we bound the information Alice and Bob can obtain if
they could use any quantum operation satisfying the P-H property.

\begin{figure}[f]
\begin{center}
\centerline{\mbox{\psfig{file=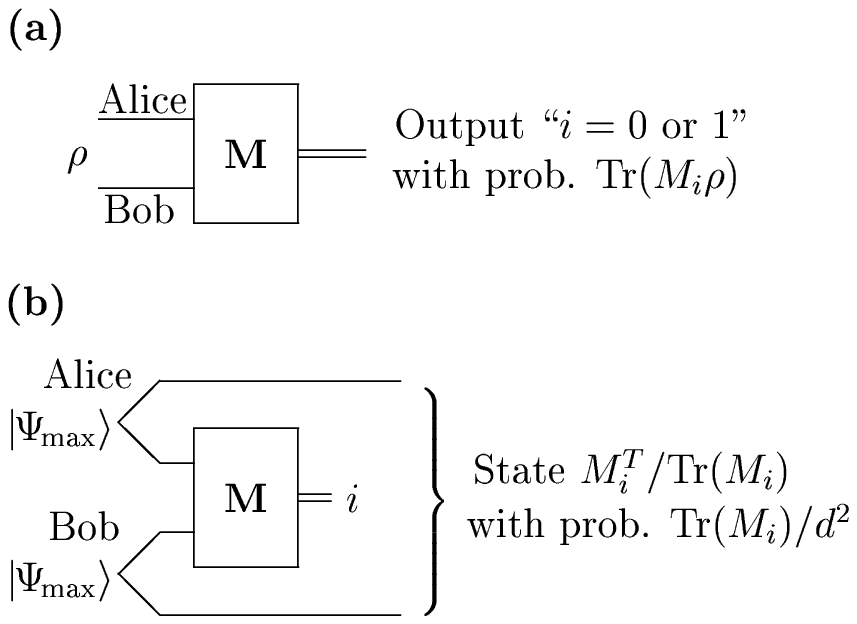,width=3in}}}
\vspace*{0.4cm}
\caption{(a) A bipartite POVM measurement with two outcomes. 
(b) Applying the bipartite POVM measurement to two halves of two 
maximally entangled states $|\Psi_{\max}\>$ results in a residual 
state which is proportional to the transpose of the POVM element 
corresponding to the measurement outcome.}
\label{fig:fig1}
\end{center}
\end{figure}

Since Alice and Bob are only interested in obtaining {\em classical}
data, we can restrict our attention to quantum operations that yield
classical outcomes only.  These operations are called POVM (Positive
Operator Valued Measure) measurements~\cite{Peresbook,Nielsen00}.
A POVM measurement is characterized by a set of positive
operators $M_i$ such that the outcome $i$ occurs with probability 
$\mbox{Tr}(M_i \rho)$.
The trace preserving condition requires that $\sum_i M_i = I$.
The set $\{ M_i \}$ is called a POVM, and each $M_i$ a {\em POVM
element}.
A POVM measurement on a bipartite input is illustrated in
Fig.~\ref{fig:fig1}(a).

Extending the discussion in Ref.~\cite{tdl:prl}, we now derive a
necessary condition for a POVM measurement to satisfy the P-H property
(and therefore, to be LOCC): each POVM element is PPT.\footnote{This
is also sufficient, see Section~\ref{sec:disc}.}

To show this, suppose Alice and Bob each create a {\em maximally
entangled state},
$|\Psi_{\max}\>=\frac{1}{\sqrt{d}}\sum_{l=0}^{d-1}|l,l\>$, in their
laboratories.  The complete state held by Alice and Bob is a product
state and is thus PPT.
Then, they apply the POVM measurement on ${\cal H}_d \otimes {\cal
H}_d$ to the two halves of the two maximally entangled states, as
shown in Fig.~\ref{fig:fig1}(b).
Suppose outcome $i$ is obtained; then the residual state in the two 
unmeasured halves is proportional to
\bea
	\rho_f \!\!\!\!& \propto &\!\!\!\!
	\sum_{l,j,m,n=0}^{d-1} 
	|l,j\>\<m,n| ~ {\rm Tr}\,[M_i \, |l,j\>\<m,n|]
\nonumber
\\
	& = & \!\!\!\! 
	\sum_{l,j,m,n=0}^{d-1} 
	\<l,j| M_i^T |m,n\> |l,j\>\<m,n| 
	=M_i^T \,,
\eea
where $M_i^T$ is the matrix transpose of $M_i$.  Thus, measurement
outcome $i$ is produced together with the state $M_i^T/$Tr$(M_i)$ in
the unmeasured halves of the maximally entangled states.  In order for
the POVM to have the P-H property, each of these states must be PPT;
this establishes that each $M_i^T$, and therefore each $M_i$, must be
PPT.

In general, we consider all possible POVMs with PPT elements.
However, if we are only interested in the probabilities of the
outcomes and our data hiding states have certain symmetries, then it
suffices to consider a class of POVMs reflecting those symmetries.

More specifically, suppose the secret $b$ is hidden in the global 
bipartite state $\rho_b$.  
Consider a POVM with PPT elements $M_i$.
The conditional probabilities of obtaining the outcome $i$ when the
secret is $b$ is given by $p_{i|b}={\rm Tr}(M_i \rho_b)$.
If $\cal T$ is a trace preserving quantum operation that is LOCC, then
${\cal T}^\dagger$ is unital (${\cal T}^\dagger[I] = I$), and the
operators ${\cal T}^\dagger [M_i]$ satisfy $\sum_i {\cal
T}^\dagger[M_i] = I$ and form another POVM with PPT elements 
(since ${\cal T}^\dagger$ satisfies the P-H property).  
Moreover, if $\cal T$ fixes all $\rho_b$ (i.e., ${\cal
T}[\rho_b]=\rho_b$), the new POVM induces conditional probabilities
$p_{i|b}' = {\rm Tr}({\cal T}^\dagger [M_i] \rho_b) = {\rm Tr}(M_i
{\cal T}[\rho_b]) = p_{i|b}$ which are equal to those induced by the
original POVM.
Hence it suffices to restrict ourselves to POVMs with elements ${\cal
T}^\dagger [M_i] \geq 0$ that are PPT and sum up to $I$. Each
quantum operation ${\cal T}$ that we will encounter reflects the
symmetries in the data hiding states, expressed by the fact that
${\cal T}$ fixes the states. The POVM-elements ${\cal T}^\dagger
[M_i]$ will possess symmetries that arise from the symmetries of
$\rho_b$, which can greatly reduce the number of independent
parameters needed to specify the POVM.  This will lead to a
significant simplification in the security analysis of our protocols.

\section{Hiding a bit in mixtures of Bell states}
\label{sec:bithid1}

In this section, we describe the basic scheme to hide a bit in a
mixture of Bell states.  This is an in-depth discussion which extends
our earlier work~\cite{tdl:prl}.  We also detail the proof of
security, and discuss both the upper and lower bounds of the
information obtained about the secret.

\subsection{The single-bit hiding scheme}

The classical bit $b = 0,1$ is hidden in the two hiding states
$\rho_0^{(n)}$ and $\rho_1^{(n)}$.  The bit should be reliably
retrievable by a quantum measurement; therefore, the states
$\rho_0^{(n)}$ and $\rho_1^{(n)}$ are required to be {\em orthogonal},
${\rm Tr}(\rho_0^{(n)}\rho_1^{(n)})=0$.  Each $\rho_b^{(n)}$ operates
on ${\cal H}_{2^n} \otimes {\cal H}_{2^n}$ and $n$ is a security
parameter.  Hence, Alice and Bob each has $n$ qubits.

$\rho_0^{(n)}$ and $\rho_1^{(n)}$ are chosen to be 
\beq
\rho_0^{(n)}=\frac{1}{|E_n|} 
\sum_{{\bf k} \in E_n} \ket{w_{{\bf k}}}\bra{w_{{\bf k}}}
\,, 
\label{r0}
\eeq
and 
\beq
\rho_1^{(n)}=\frac{1}{|O_n|} \sum_{{\bf k} \in O_n} 
\ket{w_{{\bf k}}}\bra{w_{{\bf k}}}
\,. 
\label{r1}
\eeq
Here, $\ket{w_{\bf k}}$ denotes a tensor product of $n$ Bell states
labeled by the $2n$-bit string ${\bf k}$, with the usual
identification between the four Bell states and two-bit
strings~\cite{bdsw}:
\beq
\ba{lr}
\frac{1}{\sqrt{2}}(\ket{00}+\ket{11}) \leftrightarrow 00\,, & 
\frac{1}{\sqrt{2}}(\ket{00}-\ket{11}) \leftrightarrow 01\,, \nonumber \\
\frac{1}{\sqrt{2}}(\ket{01}+\ket{10}) \leftrightarrow 10\,, & 
\frac{1}{\sqrt{2}}(\ket{01}-\ket{10}) \leftrightarrow 11\,.
\ea
\label{iden}
\eeq
In Eqs.~(\ref{r0}) and (\ref{r1}), $E_n$ is the set of $2n$-bit strings
${\bf k}$ such that the number of $11$ pairs, i.e. the number of {\em
singlet Bell states}, in ${\bf k}$, denoted as $N_{11}({\bf k})$, is
{\em even}.  $O_n$ is the set of bit strings ${\bf k}$ such that
$N_{11}({\bf k})$ is {\em odd}.
The cardinalities of $E_n$ and $O_n$, $|E_n|$ and $|O_n|$, 
satisfy the recurrence relations: 
\bea
\ba{l}
	|E_n| = |E_{n-1}||E_1| + |O_{n-1}||O_1|\,, 
\\
	|O_n| = |E_{n-1}||O_1| + |O_{n-1}||E_1|\,, 
\ea
\eea
which imply
\be
        |E_n|-|O_n| =   \left( \rule{0pt}{2.0ex} |E_{n-1}|-|O_{n-1}| \right)
                        \left( \rule{0pt}{2.0ex} |E_1|-|O_1| \right ) 
\,.
\ee
Since $|E_1|-|O_1|=3-1=2$, $|E_n|-|O_n|=2^n$ and
\beq
\ba{lr}
|E_n|=(2^{2n}+2^n)/2\,, & |O_n|=(2^{2n}-2^n)/2\,.
\ea
\eeq
%
%

If Alice and Bob can perform nonlocal measurements, then they can
simply distinguish $\rho_0^{(n)}$ from $\rho_1^{(n)}$ by measuring
along the Bell basis and counting the number of singlets.  
For example, if they share $n$ ebits, then Alice can teleport her
$n$ qubits to Bob; he then measures the $n$ pairs along the Bell
basis.

\subsection{Upper bound on the attainable information} 
\label{sec:secpf1}

A general LOCC measurement to distinguish $\rho_{0,1}^{(n)}$ is
specified by two PPT POVM elements $M_{0,1}$, both acting on ${\cal
H}_{2^n} \otimes {\cal H}_{2^n}$.
For the optimal conditional probabilities, it suffices to restrict  
to {\em Bell diagonal} POVM elements: 
\beq 
\ba{lr}
M_0=\sum_{\bf s} \alpha_{\bf s} \ket{w_{\bf s}}\bra{w_{\bf s}}\,, 
& 
M_1=\sum_{\bf s} \beta_{\bf s} \ket{w_{\bf s}}\bra{w_{\bf s}}\,.  
\ea
\label{simple}
\eeq
with $\alpha_{\bf s}, \beta_{\bf s} \geq 0$ and 
$\alpha_{\bf s} + \beta_{\bf s} = 1$ for all ${\bf s}$.  
To see this, let ${\cal T}_{\tilde{\cal P}_n}$ be the Partial
Twirl~\cite{bdsw} operation on $n$ qubits
\be
	{\cal T}_{\tilde{{\cal P}}_n}[\rho] = {1\over 2}{1 \over 4^n} 
	\sum_{P \in \tilde{{\cal P}}_n} 
	P \! \otimes \! P \, \rho \, P^\dagger \! \otimes \! P^\dagger
\,,\label{twirly}
\ee
where $\tilde{{\cal P}}_n$ is the hermitian subset of the {\em Pauli
group} ${\cal P}_n$.  The elements of ${\cal P}_n$ are tensor products
of the identity and Pauli matrices $\sig{x}{j}$, $\sig{y}{j}$ and
$\sig{z}{j}$ acting on the $j$-th qubit, with additional $\pm 1, \pm
i$ factors.
${\cal T}_{\tilde{{\cal P}}_n}={\cal T}_{\tilde{{\cal P}}_n}^\dagger$
is in the LOCC class; to implement ${\cal T}_{\tilde{{\cal P}}_n}$,
Alice and Bob agree on the same random $P$ and apply $P$ to their
respective systems.

We now show that the effect of ${\cal T}_{\tilde{{\cal P}}_n}$ is to
remove the off-diagonal elements in the Bell basis.  On one qubit,
${\cal T}_{\tilde{{\cal P}}_1}$ transforms the Bell states according to
\beq
	\sigma_{c_1 c_2} \otimes \sigma_{c_1 c_2} \ket{w_{k_1 k_2}} =
	(-1)^{k_1 \cdot c_2 \oplus k_2 \cdot c_1} \ket{w_{k_1 k_2}} 
\,.
\label{twirl}
\eeq
In this notation, $k_{1,2}$ are the two bits labeling the Bell state as
defined in Eq.~(\ref{iden}), and $c_{1,2}$ are the two bits labeling the
Pauli operators: 
%
%
$I \rightarrow \sigma_{00}$, $\sigma_x \rightarrow \sigma_{10}$,
$\sigma_z \rightarrow \sigma_{01}$, $\sigma_y \rightarrow \sigma_{11}$.
When applying ${\cal T}_{\tilde{{\cal P}}_n}$ to an arbitrary
$n$-qubit state, the phase factor in Eq.~(\ref{twirl}) assures that
all off-diagonal components in the density matrix are cancelled out
when we average over $\tilde{\cal P}_n$.
Moreover, ${\cal T}_{\tilde{{\cal P}}_n}$ fixes both $\rho_b^{(n)}$
since they are both Bell diagonal.
Following the discussion at the end of Section~\ref{sec:povm}, it
suffices to consider ${\cal T}_{\tilde{{\cal P}}_n}^\dagger[M_i]$ which are
Bell diagonal, as given by Eq.~(\ref{simple}).

\begin{figure}[f]
\begin{center}
\centerline{\mbox{\psfig{file=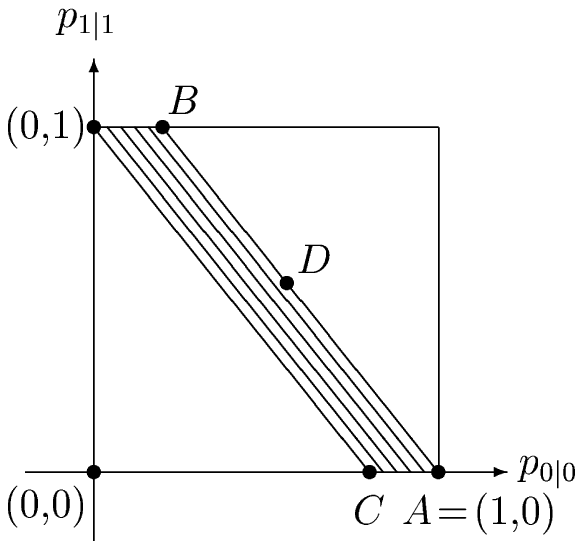,width=2.0in}}}
\vspace*{0.4cm}
\caption{Equation~(\ref{tightlt2}) restricts $(p_{0|0},p_{1|1})$ to
the above shaded region.  The points $B$, $C$, and $D$ are
respectively $({2^{-(n-1)} \over
1+2^{-n}},1)$, $({1-2^{-n} \over 1+2^{-n}},0)$, and $({1 \over 2}{2^n
+ 2 \over 2^n+1},{1 \over 2}{2^n \over 2^n-1})$.  The point $D$ is
achievable by a LOCC measurement described in
Section~\ref{tightb}. The expression $p_{0|0} + p_{1|1}$ is maximized
at $B$.  }
\label{fig:fig2}
\end{center}
\end{figure}

With the simplified form of $M_0$ and $M_1$, their partial transposes
can be evaluated directly.
Using the fact that $|w_{\bf s}\> = (\sigma_{\bf s} \otimes I) |w_{\bf 0}\>$,  
we find that  
\bea
	& & ({\bf 1} \otimes T)[M_0] 
\nonumber
\\
	& = & ({\bf 1} \otimes T) \left[\rule{0pt}{3.0ex}\right.
	\sum_{\bf s} \alpha_{\bf s} 
	(\sigma_{\bf s} \otimes I) |w_{\bf 0}\> \<w_{\bf 0}| 
	(\sigma_{\bf s} \otimes I) 
	\left.\rule{0pt}{3.0ex}\right]
\nonumber
\\
	& = & \sum_{\bf s} \alpha_{\bf s} (\sigma_{\bf s} \otimes I) 
	({\bf 1} \otimes T)[|w_{\bf 0}\> \<w_{\bf 0}|] 
	(\sigma_{\bf s} \otimes I)
\nonumber
\\
	& = & \sum_{\bf s} \alpha_{\bf s} (\sigma_{\bf s} \otimes I) 
	\left( \rule{0pt}{3.0ex} \right. \!\!
	{1 \over 2^n} \sum_{\bf k} (-1)^{N_{11}({\bf k})} 
	|w_{\bf k}\>\<w_{\bf k}| 
	\!\! \left. \rule{0pt}{3.0ex} \right) 
	(\sigma_{\bf s} \otimes I)
\nonumber
\\
	& = & {1 \over 2^n} \sum_{\bf s,k} \alpha_{\bf s} 
	(-1)^{N_{11}({\bf k})} |w_{\bf k \oplus s}\>\<w_{\bf k \oplus s}|\,. 
\eea
Hence, $({\bf 1} \otimes T)[M_0]\equiv M_0^{PT}$ is diagonal in the
Bell basis, and $M_0$ is PPT if and only if
\beq
\forall\, {\bf m}\;\; \bra{w_{{\bf m}}}M_0^{PT} \ket{w_{{\bf m}}} \geq 0
\,.
\label{expnonneg}
\eeq
Since $M_1 = I - M_0$, $M_1$ is PPT if and only if 
\beq
\forall\, {\bf m}\;\; \bra{w_{{\bf m}}}M_0^{PT} \ket{w_{{\bf m}}} \leq 1
\,.
\label{expnonlarge}
\eeq
These PPT conditions for ${\bf m}=00\ldots 0$ require 
\bea
  0 \leq \sum_{{\bf s}} \alpha_{{\bf s}} (-1)^{N_{11}({\bf s})} \leq 2^n\,,
\eea
or
\bea
  0 \leq \sum_{{\bf s} \in E_n} \alpha_{{\bf s}} 
		- \sum_{{\bf s} \in O_n} \alpha_{\bf s} \leq 2^n\,.
\label{ptcond}
\eea
We are now ready to use the expressions for $p_{0|0}$ and $p_{1|1}$:
\bea
p_{0|0} \hspace*{-2ex} & = & \hspace*{-2ex} {\rm Tr}\rho_0^{(n)}M_0=
	\frac{2}{2^{2n}+2^n}\sum_{{\bf s} \in E_n} \alpha_{\bf s}\,, 
\nonumber
\\
p_{1|1} \hspace*{-2ex} & = & \hspace*{-2ex}
	\frac{2}{2^{2n}-2^n}\sum_{{\bf s} \in O_n} \beta_{\bf s} 
	  =   \frac{2}{2^{2n}-2^n}\sum_{{\bf s} \in O_n} (1-\alpha_{\bf s})\,. 
\label{defps}
\eea
We combine Eq.~(\ref{ptcond}) with Eq.~(\ref{defps})
to obtain
\beq
 0 \leq {1 \over 2} (1+2^{-n}) p_{0|0} + {1 \over 2} (1-2^{-n}) (p_{1|1} - 1) 
 \leq 2^{-n}\,,
\label{tightlt}
\eeq
or rearranging terms, 
\beq
 {1-2^{-n} \over 2} \leq 
 {1+2^{-n} \over 2} \, p_{0|0} + {1-2^{-n} \over 2} \, p_{1|1} 
 \leq {1+2^{-n} \over 2}\,. 
\label{tightlt2}
\eeq
Equation~(\ref{tightlt2}) puts linear constraints on 
$(p_{0|0},p_{1|1})$ as depicted in Fig.~\ref{fig:fig2}, from which we find 
\beq
	|p_{0|0}+p_{1|1}-1| \leq {2^{-(n-1)} \over 1 + 2^{-n}} 
	\leq 2^{-(n-1)}
\label{finalb}
\,.
\eeq
Equation~(\ref{finalb}) implies that the measurement is not
informative when $n$ is large, since a coin flip without 
the state $\rho_b^{(n)}$ achieves $p_{0|0}+p_{1|1} = 1$. 
 
We now outline how to quantify the amount of information about the
hidden bit that can be retrieved by Alice and Bob.  
In general, Alice and Bob will have obtained, by their measurements
and operations, a {\em multistate} outcome from which the final 
outcome is {\em inferred}.  This corresponds to the following process
\beq
        B \stackrel{{\cal M}}{\rightarrow} 
        Y \stackrel{{\cal D}}{\rightarrow} \hat{B}\,,
\label{proc}
\eeq
where $Y$ is a multistate random variable representing the outcome of
the general LOCC measurement ${\cal M}$ and ${\cal D}$ is a decoding
scheme to infer the bit $b$ from $Y$. 
In principle, $Y$ can contain more information about the hidden bit
than the final inferred outcome, since information can be lost when
decoding from a multistate random variable to a binary one.
In Appendix~\ref{sec:mutualinfo}, we show that any multistate random
variable $Y$ obtained in a scheme such as in Eq.~(\ref{proc}) under
the condition of Eq.~(\ref{finalb}) conveys at most $H(B)/2^{n-1}$
bits of information on $B$.  
Here $H(B)$ is the Shannon information of the hidden bit.
In other words, $I(B:Y) \leq H(B)/2^{n-1}$, and only a vanishing
fraction of the Shannon information of the hidden bit can be obtained.

We have given an elementary proof of Eq.~(\ref{finalb}).  We now give
an alternative proof that is more easily generalized to hide multiple 
bits.  This proof uses the fact that $\rho_b^{(n)}$ are two
extremal {\em Werner states}~\cite{werner:lhv}:
\bea
	\rho_0^{(n)} & = & \frac{1}{2^n(2^n+1)} (I+2^n H_n) \,, 
\label{rec0}
\\
	\rho_1^{(n)} & = & \frac{1}{2^n(2^n-1)} (I-2^n H_n) \,,
\label{rec1}
\eea
where
\bea
	H_n = \left( ({\bf 1} \otimes T)[\ket{\Phi^+}\bra{\Phi^+}] 
						\right)^{\otimes n}
	=  \frac{1}{2}{1 \over 4^n} 
	\sum_{P \in \tilde{\cal P}_n} P \otimes P\,, 
\label{hn}
\eea
with $P$ ranging over the hermitian subset $\tilde{\cal P}_n$ of
${\cal P}_n$.  (Equation~(\ref{hn}) will be proved in
Section~\ref{sec:prepstate}, see Eq.~(\ref{lasth}).)
Equations~(\ref{rec0}) and (\ref{rec1}) can be proved by induction using
Eq.~(\ref{hn}) and the recursive expressions of $\rho_0^{(n)}$ and
$\rho_1^{(n)}$:
\bea
\rho_0^{(n)} = q_n ~\rho_1^{(n-1)}\otimes \rho_1^{(1)}+
(1-q_n) ~\rho_0^{(n-1)} \otimes \rho_0^{(1)}\,,
\nonumber
\\
\rho_1^{(n)} = p_n ~\rho_0^{(n-1)}\otimes \rho_1^{(1)}+
(1-p_n) ~\rho_1^{(n-1)} \otimes \rho_0^{(1)}\,,
\label{recur2}
\eea
where 
\beq
q_n={2^{n-1}-1\over 2(2^n+1)} 
\,,
\ \ \ 
p_n={2^{n-1}+1\over 2(2^n-1)}
\,.
\label{recur3}
\eeq
It is known that $I$ and $H_n$, and therefore $\rho_b^{(n)}$, are
invariant under the bilateral action $U \otimes U$ for any unitary
operation $U \in U(2^n)$.  
Hence $\rho_b^{(n)}$ are fixed by the {\em Full Twirl} operation: 
\beq
	{\cal T}_{U(2^n)}[\rho] = \frac{1}{{\rm Vol}(U)} \int \; 
	dU (U \otimes U)~ \rho ~(U^{\dagger} \otimes U^{\dagger})
\,.
\eeq
Like the partial twirl ${\cal T}_{\tilde{{\cal P}}_n}$, the Full Twirl
is also self-adjoint and in the LOCC class.
Moreover, it is also known that the effect of the Full Twirl is to
turn any operator into a linear combination of $I$ and 
$H_n$~\cite{werner:lhv}.
Following the discussion in Section~\ref{sec:povm}, when considering the
conditional probabilities, $M_0$ and $M_1$ can be taken to be
\bea
	M_0 & = & \alpha I + \beta ~2^n H_n\,,
\nonumber
\\	M_1 & = & (1- \alpha) I - \beta ~2^n H_n
\label{werner2}
\,.
\eea
Using 
\be
	{\rm Tr}(H_n) = 1 
\,, 
	\ \ \ {\rm Tr}(H_n^2) = 1
\,,
\label{traceh}
\ee
we find that
\bea
	p_{0|0} & = & \alpha + \beta\,,
\nonumber
\\	
	p_{1|1} & = & (1-\alpha) + \beta\,,
\eea
and
\be
	p_{0|0} + p_{1|1} = 1 + 2 \beta\,.
\ee
A bound on the above expression can be found by extremizing the value
of $\beta$ subject to the constraint that $M_{0,1}$ in
Eq.~(\ref{werner2}) are positive and PPT.  From
Refs.~\cite{nptnond1,nptnond2} we know that an operator $a I + b 2^n
H_n$ has nonnegative eigenvalues and is PPT if $-\frac{a}{2^n} \leq b
\leq a$.  Applying the constraints to Eq.~(\ref{werner2}), we have
%
%
\be 
	\alpha-1\leq \beta \leq \alpha \,, \ \ \ 
	-\frac{\alpha}{2^n} \leq \beta \leq \frac{1-\alpha}{2^n}\,.
\ee
Eliminating $\alpha$ from the above inequalities, we obtain 
$|\beta| \leq {1 \over 2^n +1}$ and thus
\be
	|p_{0|0} + p_{1|1}-1| \leq {2 \over 2^n + 1}\,,\label{cleanbound}
\ee
which is what we set out to prove.

\subsection{A tight LOCC measurement scheme} 
\label{tightb}

We can give a lower bound on the attainable value of $p_{0|0} +
p_{1|1} - 1$ by analyzing a particular LOCC measurement scheme to
distinguish $\rho_0^{(n)}$ from $\rho_1^{(n)}$.  In this scheme, Alice
and Bob try their best to distinguish whether each Bell pair is a
singlet state or not; then they take the parity of all the results.
The pairwise strategy is for Alice and Bob to measure their qubits in
the $\{ \ket{0}, \ket{1}\}$ basis, and to infer a singlet whenever the
results disagree.  For each pair, this gives conditional probabilities
\begin{equation}
	p_{0|0}^{(1)}={2 \over 3}\,,\ \ \ p_{1|1}^{(1)}=1\,.
\end{equation}
Using Eqs.~(\ref{recur2}) and (\ref{recur3}), 
we can immediately write the conditional probabilities of interest
for the $n$-pair measurement:
\bea
p_{0|0}^{(n)} & = & (1-q_n) \left[ p_{0|0}^{(1)} ~ p_{0|0}^{(n-1)} 
		+ p_{1|0}^{(1)} ~ p_{1|0}^{(n-1)} \right] 
\nonumber
\\
	  &   & + ~q_n~ \left[p_{0|1}^{(1)} ~ p_{0|1}^{(n-1)} 
		+ p_{1|1}^{(1)} ~ p_{1|1}^{(n-1)} \right]\,,
\nonumber
\\
p_{1|1}^{(n)} & = & (1-p_n) \left[p_{0|0}^{(1)} ~ p_{1|1}^{(n-1)} 
		+ p_{1|0}^{(1)} ~ p_{0|1}^{(n-1)} \right]
\nonumber
\\
	  &   & + ~p_n~ \left[p_{0|1}^{(1)} ~ p_{1|0}^{(n-1)} 
		+ p_{1|1}^{(1)} ~ p_{0|0}^{(n-1)} \right]\,.
\eea
It is easy to confirm that these expressions are satisfied by
\begin{eqnarray}
	p_{0|0}^{(n)} & = & {1 \over 2}{2^n + 2 \over 2^n+1}
\,,
\nonumber
\\
	p_{1|1}^{(n)} & = & {1 \over 2}{2^n \over 2^n-1}\,.
\end{eqnarray}
This set of $(p_{0|0}^{(n)},p_{1|1}^{(n)})$ is plotted in
Fig.~\ref{fig:fig2} as point $D$. 
Note that, for all $n$, they saturate the last inequality of
Eq.~(\ref{tightlt2}) and therefore our security result is tight for
this value of $(p_{0|0},p_{1|1})$.  Because of convexity, this
gives a tight result along the full line segment connecting point $D$
with the point $A=(1,0)$.

\section{General lower bound for hiding a single bit}
\label{sec:genlower}

Now we show that some information can always be 
extracted when orthogonal hiding states $\rho_{0,1}$ are used.  
The intuitive reason is that {\em state tomography} can be performed
by LOCC.  Given a large number of copies of $\rho_{b}$, it is possible
to identify $\rho_b$ and thus $b$.  Therefore, each copy {\em must}
carry a non-zero amount of information.
The precise statement is the following: 
\medskip

\begin{theo} 
For all pairs $\rho_{0,1}$ on ${\cal H}_{2^n} \otimes {\cal H}_{2^n}$
such that $\mbox{Tr} (\rho_0 \rho_1) = 0$, there exists a two-outcome
LOCC measurement such that
\begin{equation}
	p_{0|0} + p_{1|1} - 1 \geq {1 \over \sqrt{16^n-1}}
	\sqrt{1+(p_{0|0}-p_{1|1})^2 } \,.
\label{curve}
\end{equation}
It is immediate that  
\begin{equation}
	p_{0|0}+p_{1|1}-1 \geq {1 \over \sqrt{16^n - 1}} \,.
\label{corr}
\end{equation}
\label{theogenlower}
\end{theo}

\begin{figure}[f]
\begin{center}
\centerline{\mbox{\psfig{file=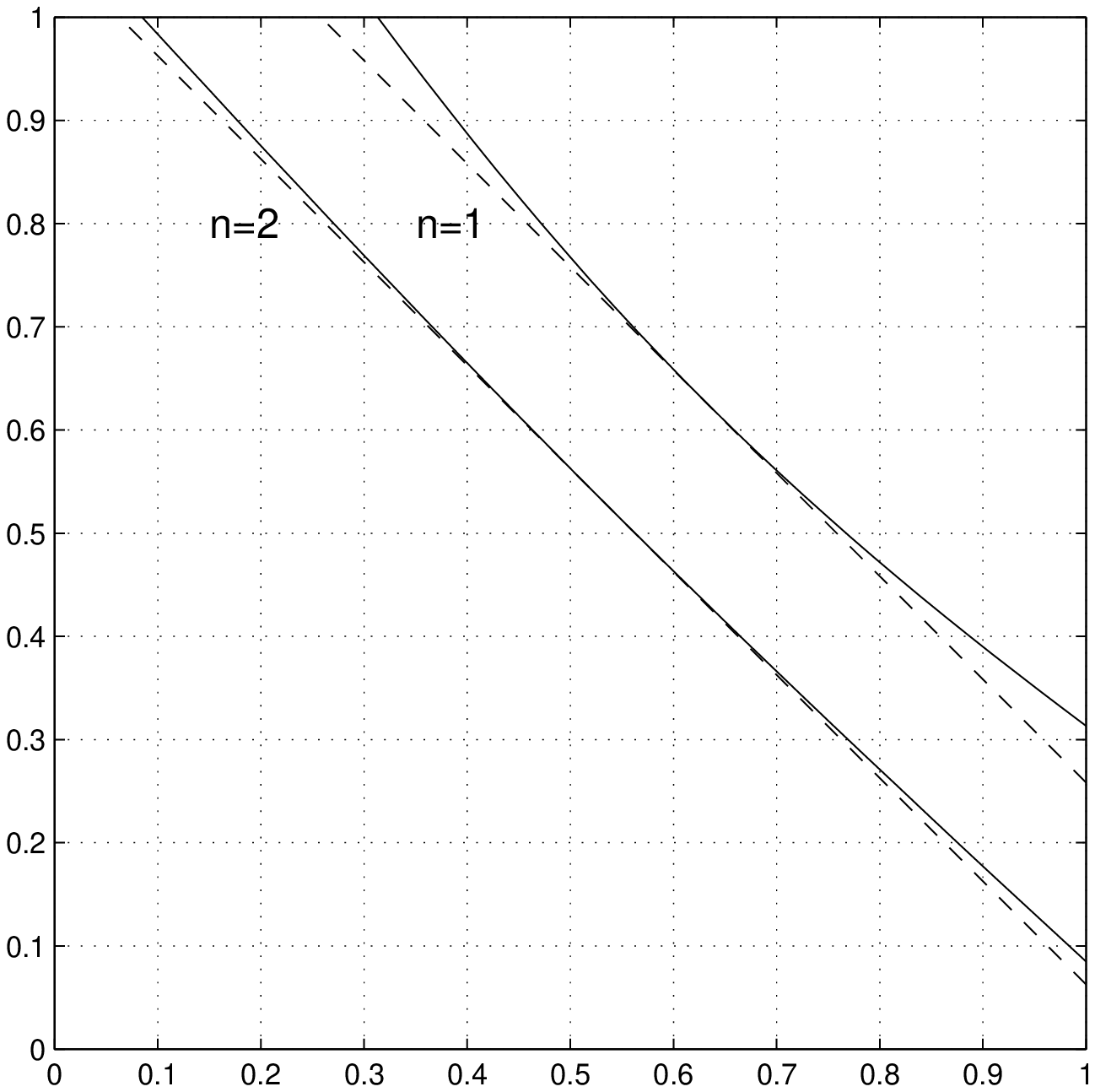,width=3in}}}
\vspace*{0.4cm}
\caption{For any pair of orthogonal states $\rho_{0,1}$ on ${\cal
H}_{2^n} \otimes {\cal H}_{2^n}$, there exists an LOCC POVM with
probabilities $p_{0|0}$, $p_{1|1}$ above the curves shown for $n=1$
and $n=2$.  The dashed lines are the simpler, weaker bound of
Eq.~(\protect\ref{corr}).}
\label{fig:boundcurve}
\end{center}
\end{figure}

These bounds are plotted for $n=1$ and $n=2$ in Fig. \ref{fig:boundcurve}.

In Appendix~\ref{prove}, we give a proof of this Theorem using a
Lagrange multiplier analysis, together with an alternative simple
proof for the weaker result Eq.~(\ref{corr}).  
This bound is not necessarily tight; the proof of Theorem
\ref{theogenlower} gives no indication of whether or not the right
hand side of Eq.~(\ref{curve}) could be larger.  However, its
approximate behavior, i.e. $p_{0|0}+p_{1|1}-1 \geq \frac{1}{2^{O(n)}}$
is in accordance with our findings in Section~\ref{sec:bithid1}.

\section{Hiding Multiple Bits}
\label{sec:multiple}

In this section we will prove the security of hiding multiple bits
bitwise with the scheme described in Section~\ref{sec:bithid1}.
Let ${\bf b} = (b_1,b_2,\cdots,b_k)$ be a $k$-bit string to be hidden.  
The hiding state is 
\be
	\rho_{\bf b}^{(n)} = \bigotimes_{l=1}^k \rho_{b_l}^{(n)}\,.  
\ee
Showing the security of the bitwise scheme is nontrivial.
First, one of the hiding states $\rho_1^{(n)}$ is entangled (see
Section~\ref{sec:prepstate}), and the entanglement in part of the
system may help decode the partial secret in the rest of the system.
Second, joint measurement on all $k$ tensor product components may
provide more information than a measurement on each component
separately.
The security of multiple-bit hiding is established using the symmetry
of $\rho_{0}^{(n)}$ and $\rho_{1}^{(n)}$, as captured by their
Werner-state representation (Eqs.~(\ref{rec0}) and (\ref{rec1})).

In the setup for hiding $k$ bits, each POVM element $M_{\bf i}$ has a
$k$-bit index ${\bf i}$ and acts on ${\cal H}_{2^{n k}} \otimes {\cal
H}_{2^{n k}}$.
Since $\rho_{\bf b}^{(n)}$ is invariant under 
$({\cal T}_{U(2^n)})^{\otimes k}$, $M_{\bf i}$ can be parametrized as
\beq
	M_{\bf i} = \sum_{p=0}^k \sum_{{\bf m}: w_h({\bf m})=k-p} 
		\alpha_{p,{\bf m}}^{\bf i} X_{\bf m}\,.\label{expnd}
\eeq
where $X_{\bf m} = \bigotimes_{l=1}^k H_n^{m_l}$ and $w_h({\bf m})$ is
the Hamming weight of the $k$-bit string ${\bf m}$.
Here $m_l$ denotes the $l$th bit of ${\bf m}$, $H_n^0=I$ and
$H_n^1=H_n$.  The number of $H_n$ in $X_{\bf m}$ is thus $w_h({\bf
m})$.  With this parametrization the trace preserving condition
$\sum_{\bf i} M_{\bf i}=I$ implies that
\beq
\ba{lr}
	\sum_{\bf i} \alpha^{\bf i}_{k,{\bf m} = {\bf 0}} = 1\,, &
%
	\forall {\bf m} \neq {\bf 0} \; 
	\sum_{{\bf i}} \alpha_{p,{\bf m}}^{\bf i}=0 \,.
\ea
\label{2cons}
\eeq
Next we consider the constraint $({\bf 1}\otimes T)[M_{\bf i}] \geq
0$. The partial transpose replaces the operator $H_n$ by
$P_+=(\ket{\Phi^+}\bra{\Phi^+})^{\otimes n}$. Therefore we have
\beq
	({\bf 1} \otimes T)[M_{\bf i}] = 
	\sum_{p=0}^k \sum_{{\bf m}: w_h({\bf m})=k-p} 
	A_{p,{\bf m}}^{\bf i} Y_{\bf m}\,,
\label{mi}
\eeq
where $Y_{\bf m}$ is a tensor product of the orthogonal projectors
${\bf 1}-P_+$ and $P_+$, such that $P_+$ occurs where the $k$-bit
string ${\bf m}$ has a $1$. Here the coefficients $A_{p,{\bf m}}^{\bf
i}$ are particular sums of the coefficients $\alpha_{p,{\bf m}}^{\bf
i}$ of Eq.~(\ref{expnd}):
\beq
	A_{p,{\bf m}}^{\bf i} = \sum_{p \leq l \leq k} 
	\sum_{\stackrel{{\bf n}: w_h({\bf n}) = k - l}
			{
\wedge_i({\bar n}_i\vee m_i)=1
}
} 
	\alpha_{l,{\bf n}}^{\bf i}\,.
\eeq
The Boolean-logic condition on ${\bf n}$ in the summation,
$\wedge_i({\bar n}_i\vee m_i)=1$, can be expressed in ordinary
language by saying that the string {\bf n} must have 0s wherever the
sting {\bf m} has 0s.  A necessary (and, in fact, sufficient)
condition for the positivity of Eq.~(\ref{mi}) is that these
coefficients $A_{p,{\bf m}}^{\bf i} \geq 0$. When $p=k$, this implies
that
\beq
	A_{k,{\bf m}={\bf 0}}^{\bf i} = 
	\alpha_{k,{\bf 0}}^{\bf i} \geq 0 \,. 
\label{init}
\eeq
Together with Eq.~(\ref{2cons}) we obtain for all ${\bf i}$
\beq
	0 \leq \alpha_{k,{\bf 0}}^{\bf i} \leq 1 \,.
\label{lp1}
\eeq
We will bound the coefficients $\alpha_{p,{\bf n}}^{\bf i}$ for all
${\bf i}$, $p$ and ${\bf n}$. These upper and lower bounds,
\beq
	L_p \leq \alpha_{p,{\bf n}}^{\bf i} \leq U_p \,, 
\eeq
will be obtained recursively as we decrease $p$ from $k$.  The
recurrence starts with $L_k=0$ and $U_k=1$, Eq.~(\ref{lp1}).  Let us
assume that we can determine $L_p$ for $p<k$. Then $U_p$ directly
follows, using the second equation in Eq.~(\ref{2cons}): we have
\beq
	\alpha_{p,{\bf n}}^{{\bf j}} = 
-\sum_{{\bf i} \neq {\bf j}}\alpha_{p,{\bf n}}^{\bf i} \leq -(2^k-1)L_p\,, 
\eeq
or $U_p=-(2^k-1)L_p$ for $p<k$. Then we need to determine $L_p$, which
can be done using the PPT condition. We can express $A_{p,{\bf
m}}^{\bf i} \geq 0$ as
\bea
	\alpha_{p,{\bf m}}^{\bf i} & \geq & -\sum_{p < l \leq k} 
	\sum_{\stackrel{{\bf n}: w_h({\bf n})=k-l}
			{\wedge_i(\bar{n}_i\vee m_i)=1}} 
 	\alpha_{l,{\bf n}}^{\bf i} 
\nonumber
\\
	& \geq & - \sum_{p < l \leq k} U_l {k-p \choose l-p} \,.
\eea
Or, for $p<k$ 
\beq
	L_p = - \sum_{p < l \leq k} U_l {k-p \choose l-p} \,,
\label{recl}
\eeq
so, in terms of $L_l$:
\beq
	L_p = -1+(2^k-1)\sum_{p < l < k} L_l {k-p \choose l-p} \,.
\eeq
This recursion can be solved, giving
\beq
	L_p = -\sum_{l=1}^{k-p}(1-2^k)^{l-1} \sum_{j=1}^l
	(-1)^j {l\choose j} j^{k-p} \,,
\eeq
which can be rewritten in terms of the Sterling numbers of the 
second type, $\{{x \atop y}\}$, as
\beq
	L_p = -\sum_{l=1}^{k-p} (2^k-1)^{l-1} l! 
	\left\{{k-p \atop l}\right\} 
\,.
\label{ster}
\eeq

Let us consider the probabilities 
$p_{{\bf i}|{\bf b}} = {\rm Tr}\, (M_{\bf i}\rho_{\bf b})$. 
Using the fact that ${\rm Tr}\, H_n={\rm Tr}\, H_n^2=1$, 
it can be shown that for ${\bf m}$ with Hamming weight $k-p$
\beq
	  {\rm Tr}\;( X_{\bf m} \rho_{\bf b}) 
	= \frac{(-1)^{{\bf b} \cdot {\bf m}}}{2^{n(k-p)}}\,. 
\eeq
Thus 
\beq
	p_{{\bf i}|{\bf b}} = \sum_{p=0}^k \sum_{{\bf m}: w_h({\bf m})=k-p} 
	\alpha_{p,{\bf m}}^{\bf i} 
	\frac{(-1)^{{\bf b} \cdot {\bf m}}}{2^{n(k-p)}}\,,
\eeq
or
\beq
	p_{{\bf i}|{\bf b}} = \alpha_{k,{\bf 0}}^{\bf i} + 
	\sum_{p=0}^{k-1} \sum_{{\bf m}: w_h({\bf m})=k-p} 
	\alpha_{p,{\bf m}}^{\bf i} 
	\frac{(-1)^{{\bf b} \cdot {\bf m}}}{2^{n(k-p)}}\,.
\eeq
We bound the magnitude of the last term as 
\bea
	& & \left| \sum_{p=0}^{k-1} \sum_{{\bf m}: w_h({\bf m})=k-p} 
	\alpha_{p,{\bf m}}^{\bf i} 
	\frac{(-1)^{{\bf b} \cdot {\bf m}}}{2^{n(k-p)}} \right|  
\nonumber
\\	& & \hspace*{3ex} \leq \Delta =  
	(2^k-1) \sum_{p=0}^{k-1} {k \choose k-p} \frac{|L_p|}{2^{n(k-p)}}\,,
\label{bou}
\eea
so we bound the conditional probabilities
\beq
		\alpha_{k,{\bf 0}}^{\bf i} - \Delta 
	\leq 	p_{{\bf i}|{\bf b}}
	\leq	\alpha_{k,{\bf 0}}^{\bf i} + \Delta\,.
\label{bbd}
\eeq
$\Delta$ enters the bound for the mutual information obtainable by
Alice and Bob. Ideally we would like to use Eq.~(\ref{bbd}) to bound
the retrievable attainable mutual information assuming an arbitrary
probability distribution for the hiding states $\rho_{{\bf b}}^{(n)}$,
as we did in the case of hiding a single bit.  We do not know if the
proof technique of Theorem \ref{intmutual} in
Appendix~\ref{sec:mutualinfo} is applicable in the multiple-bit case,
so we will have recourse to another method which provides a bound in
the case of equal probabilities $p_{\bf b}=\frac{1}{2^k}$.

We first bound the total error probability
\be
	p_e = 1 - \sum_{\bf b} p_{\bf b} \, p_{{\bf b}|{\bf b}} 
	    = 1 - {1 \over 2^k} \sum_{\bf b} p_{{\bf b}|{\bf b}}\,.
\ee
Using Eq.~(\ref{bbd}) we get
\beq
1-{1\over2^k}\sum_{\bf b}(\alpha_{k,{\bf 0}}^{\bf b}+\Delta)\leq p_e\leq
1-{1\over2^k}\sum_{\bf b}(\alpha_{k,{\bf 0}}^{\bf b}-\Delta)
\,,
\eeq
and, with Eq.~(\ref{2cons}), we obtain
\beq
		1 - {1 \over 2^k} - \Delta 
	\leq  	p_e
	\leq 	1 - {1 \over 2^k} + \Delta
\,.
\label{bracket}
\eeq

\begin{figure}
\begin{center}
\epsfxsize=8.5cm 
\epsffile{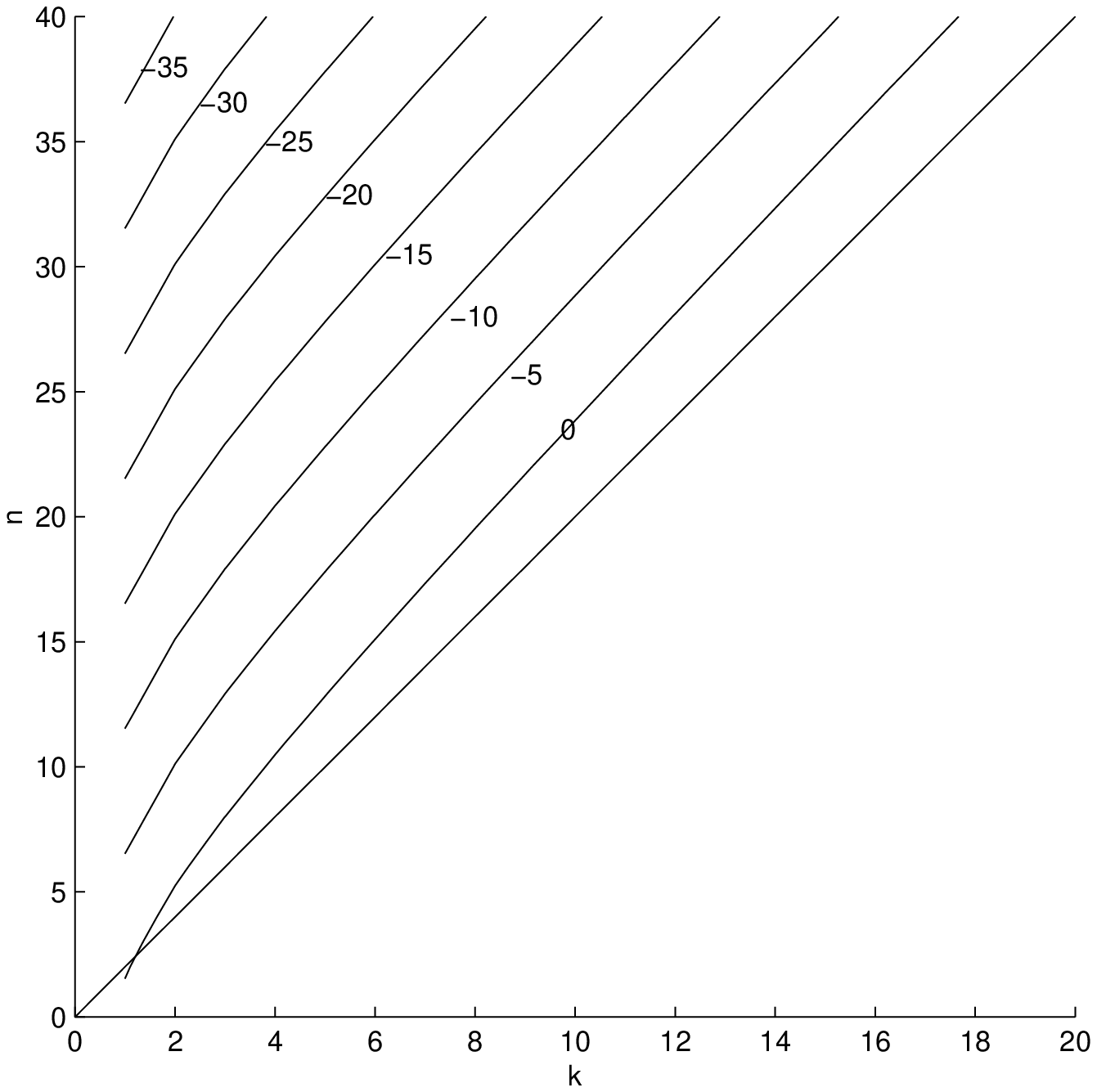}
\caption{Contours of constant upper bound on $\log I({\bf B}:Y)$,
Eq.~(\ref{ibou}), while varying $n$ (the vertical axis) and $k$ (the
horizontal axis). The bound on $\log I({\bf B}:Y)$ is calculated using
the expression for $\Delta$ in Eq.~(\ref{bou}) with $L_p$ in
Eq.~(\ref{ster}).}
\label{mbound}
\end{center}
\end{figure}

With this bound on the error probability it is possible to bound the
mutual information between the $k$ hidden bits ${\bf b}$ and a
multi-outcome measurement by Alice and Bob, similarly to the single
bit case, see Eq.~(\ref{proc}). For a process such as Eq.~(\ref{proc})
where $B$ and $\hat{B}$ are replaced by $k$ bit random variables ${\bf
B}$ and $\hat{{\bf B}}$, it can be shown\footnote{This was proved by
V. Castelli, October 2000.} that the mutual information $I({\bf B}:Y)
\leq H({\bf B})+\log(1-p_e)$.  Using Eq.~(\ref{bracket}), this implies
that
\beq
	I({\bf B}:Y) \leq {2^k\over\ln 2} \Delta
\,.
\label{ibou}
\eeq
Figure~\ref{mbound} shows an exact calculation of this bound as a
function of $n$ and $k$.  We can show that these bound curves have a
simple form in the region of interest by a further examination of
Eqs.~(\ref{ster}) and (\ref{bou}).  It is straightforward to
demonstrate that, if $k \gg 1$, Eq.~(\ref{ster}) is dominated by its
final term, so that we can approximate
\beq
	L_p\approx-(2^k-1)^{k-p-1}(k-p)!\,.
\eeq
With this we can write Eq.~(\ref{bou}) as
\bea
	\Delta & \approx & \sum_{s=1}^k {k \choose s}
		{(2^k-1)^s \over 2^{ns}} s! 
\nonumber
\\
	& = & k! \sum_{i=0}^{k-1} 
	\left({2^k-1 \over 2^n} \right)^{k-i} {1 \over i!}\,.
\eea
It is easy to show that if $2^{n-k} \gg k$ then the last term in this
sum dominates, so we obtain
\beq
	\Delta \approx k 2^{k-n}
\,,
\eeq
and Eq.~(\ref{ibou}) becomes
\beq
	I({\bf B}:Y) \stackrel{\textstyle <}{\sim} {k 2^{2k-n} \over \ln 2}
\,.
\label{fini}
\eeq
The curves shown in Fig.~\ref{mbound} are excellently approximated by
this expression.

This result shows that if we fix $k$, there always exists an $n$ large
enough so that the information that Alice and Bob can gain is
arbitrarily small.  
Equation~(\ref{fini}) says that, for a given security parameter
$I({\bf B}:Y) \leq \epsilon$, $n$ should grow, in the large $k$ limit,
as
\beq
	n(k) \rightarrow 2 k + \log k + \log \log {\rm e} + \log(1/\epsilon) 
\,.
\eeq
It is interesting to note that the above bound is much weaker than in
the case when information is additive over the different bits hidden,
which would imply
\beq
	n(k) \rightarrow \log k + \log(1/\epsilon) 
\,.
\eeq
Sufficient conditions for information to be additive will be
discussed in Section~\ref{sec:sepadd}.

\section{Preparation of the hiding states}
\label{sec:prepstate}

We consider the question of how the hider can efficiently produce the
states $\rho_0^{(n)}$ and $\rho_1^{(n)}$ starting with minimal
entanglement between the two shares. The defining representation of the
states, Eqs.~(\ref{r0}) and (\ref{r1}), suggests a method to create
the hiding states by picking $n$ Bell states with the correct number
of singlets (even or odd).  This method is computationally efficient, but
uses a lot of quantum entanglement between the shares, namely $n$
ebits per hiding state.
The alternative representation of $\rho_0^{(n)}$ and $\rho_1^{(n)}$ as
Werner states, Eqs.~(\ref{rec0}) and (\ref{rec1}), can be used to show
that they have $0$ and $1$ ebit of entanglement of formation
respectively~\cite{nptnond1,vollwerner}.
We first describe an efficient LOCC preparation for $\rho_0^{(n)}$
from the unentangled pure state $|{\bf 0}\>=|0\>^{\otimes n}\otimes
|0\>^{\otimes n}$. By efficient we mean that the number of quantum and
classical computational steps scales as a polynomial in $n$, the
number of qubits in each share.
Then, using the recursive relations Eq.~(\ref{recur2}),
and the preparation for $\rho_0^{(n)}$, we give a preparation for
$\rho_1^{(n)}$ using exactly $1$ ebit for arbitrary $n$. Note that this 
is a tight construction in terms of entanglement, since $\rho_1^{(n)}$ has
an entanglement of formation of 1 ebit.

\subsection{Preparing $\rho_0^{(n)}$}

Recall from Section~\ref{sec:bithid1} that the Full Twirl 
\beq
	{\cal T}_{U(2^n)}[\rho]=\frac{1}{\mbox{Vol}(U)} \int \, 
	dU (U \otimes U)\, \rho \,(U^{\dagger} \otimes U^{\dagger})\,, 
\eeq
has only two linearly independent invariants $I$ and $H_n$, and it transforms
any state to a Werner state.  
Moreover, note that 
\be
	{\rm Tr}({\cal T}_{U(2^n)}[\rho] H_n)
	= {\rm Tr}(\rho {\cal T}_{U(2^n)}^\dagger[H_n])
	= {\rm Tr}(\rho H_n)
\,.\label{hinvar}
\ee
Hence, the Full Twirl transforms $\rho$ to a Werner state with the 
same overlap with $H_n$.  
{From} Eqs.~(\ref{rec0}) and (\ref{traceh}), 
${\rm Tr} \left( \rho_0^{(n)} H_n \right) = \frac{1}{2^n}$; 
this also equals ${\rm Tr} \left( |{\bf 0}\>\<{\bf 0}| H_n \right)$
using Eq.~(\ref{hn}).
Hence,  
\be
	\rho_0^{(n)} = {\cal T}_{U(2^n)}\left[ |{\bf 0}\>\<{\bf 0}| \right]
\,, 
\ee 
and $\rho_0^{(n)}$ could be prepared by applying ${\cal T}_{U(2^n)}$ to 
$|{\bf 0}\>$. 

The Full Twirl, interpreted as the application of a random bilateral
unitary, is not efficient to implement.
Applying a unitary transformation selected at random is a hard
problem: almost all unitary transformations take an exponential time
in the number of qubits to accurately approximate~\cite{knill_approx}.
In the following, we give an efficient implementation of the Full
Twirl, first by showing that it is {\em equivalent} to randomizing
over the {\em Clifford group} only, and second by providing methods 
to select and implement a random Clifford group element efficiently.

\subsubsection{Clifford Twirl}

The Clifford group ${\cal C}$ has appeared in quantum information
theory as an important group in the context of quantum error
correcting codes~\cite{calderetal,gotthesis}. The Clifford group
${\cal C}_n \subset U(2^n)$ is the normalizer of the Pauli group
${\cal P}_n$.
The order of the Clifford group acting on $n$ qubits is 
$|{\cal C}_n|=2^{n^2+2n+3}\Pi_{j=1}^n (4^j-1)$~\cite{calderetal}. 

It is useful to consider how each $c \in {\cal C}_n$ acts on ${\cal
P}_n$ by conjugation.
As the conjugation map is reversible, ${\cal C}_n$ is a subgroup 
of the permutation group acting on ${\cal P}_n$.
Each $c$ is specified by the $2n$ images of the generators of ${\cal
P}_n$ $a_i = c \sig{x}{i} c^\dagger$ and $b_i = c \sig{z}{i}
c^\dagger$ for $i = 1,\cdots,n$.
There are restrictions on these images, since conjugation preserves
the eigenvalues (and thus the trace), the commutation relations, and
the multiplicative structure of the Pauli group.
Because of eigenvalue preservation, conjugation preserves the
hermitian subset $\tilde{\cal P}_n$ (introduced earlier, see
Eq.~(\ref{twirly})).
The images are $2n$ traceless hermitian Pauli operators satisfying the
commutation relations $\forall i,j$ $[a_i,a_j] = [b_i,b_j] = 0$,
$\forall i \neq j$ $[a_i,b_j] = 0$ and $\{a_i,b_i\} = 0$.
Each $c \in {\cal C}_n$ can be explicitly constructed: first choose
$a_1 \neq I$, then choose $a_2$ to commute with $a_1$, and $a_2 \notin
\{I,a_1\}$, then choose $a_3$ to commute with $a_1,a_2$ and $a_3
\notin \{I,a_1,a_2,a_1 a_2\}$, and so on until $a_n$ is chosen.  Each
$b_i$ can be chosen to anticommute with $a_i$ and commute with all
other $a_j$ and $b_1,\cdots,b_{i-1}$.
If $(a_1,\cdots,a_n,b_1,\cdots,b_n)$ is a valid set of images
corresponding to a Clifford group element $c$, 
$((-1)^{\epsilon_1} a_1,\cdots,(-1)^{\epsilon_n} a_n, 
(-1)^{\epsilon_{n+1}} b_1,\cdots,(-1)^{\epsilon_{2n}} b_n)$
is another valid set corresponding to
\be
	c\, (\sigma_z^{\epsilon_1} \sigma_x^{\epsilon_{n+1}} \otimes 
	\ldots \otimes  
	\sigma_z^{\epsilon_n} \sigma_x^{\epsilon_{2n}})
\,,
\label{switch}
\ee
where each $\epsilon_i=0,1$. 

We define the ``Clifford twirl'' on ${\cal H}_{2^n} \otimes {\cal
H}_{2^n}$ to be the operation
\beq
	{\cal T}_{{\cal C}_n}[\rho] = 
	\frac{1}{|{\cal C}_n|} \sum_{c \in {\cal C}_n} 
	(c \otimes c) \,\,\rho\,\, (c^{\dagger} \otimes c^{\dagger})
\,.
\label{ct}
\eeq
We now prove that ${\cal T}_{{\cal C}_n}={\cal T}_{U(2^n)}$.  We need
to show that these two quantum operations transform any state to the
same output.
It suffices to show that 
(1) ${\cal T}_{{\cal C}_n} = {\cal T}_{{\cal C}_n}^\dagger$, 
(2) ${\cal T}_{{\cal C}_n}[I]=I$, 
(3) ${\cal T}_{{\cal C}_n}[H_n]=H_n$, and 
(4) ${\cal T}_{{\cal C}_n}$ transforms any state to a linear combination 
    of $I$ and $H_n$.  
Conditions (1)-(4) ensure that ${\cal T}_{{\cal C}_n}$ transforms any
state to a Werner state with the correct overlap with $H_n$, i.e.,
${\rm Tr}\, (H_n \rho)={\rm Tr}\, (H_n {\cal T}_{{\cal C}_n}[\rho])$
(cf. Eq.~(\ref{hinvar})).
Conditions (1) and (2) are obvious.  Condition (3) can be proved by writing 
\bea
	H_n & = & 
	\left(({\bf 1}\otimes T)[|\Phi^+\>\<\Phi^+|] \right)^{\otimes n}
\nonumber
\\
	& = & {1 \over 4^n} (I \otimes I + \sigma_x \otimes \sigma_x + 
	\sigma_y \otimes \sigma_y + \sigma_z \otimes \sigma_z)^{\otimes n}
\nonumber
\\	& = & {1\over 2}{1 \over 4^n} \sum_{P \in \tilde{\cal P}_n} P \otimes P
\,.
\label{lasth}
\eea
Since conjugation by each $c \in {\cal C}_n$ only permutes the terms
in this sum over $\tilde{\cal P}_n$, $H_n$ is invariant under ${\cal
T}_{{\cal C}_n}$.  This establishes (3).

To show condition (4), we consider the action of ${\cal T}_{{\cal
C}_n}$ on a basis for the density matrices.  We choose the basis 
$P_1 \otimes P_2 \in \tilde{\cal P}_{n} \otimes \tilde{\cal P}_{n}$.
We already know that ${\cal T}_{{\cal C}_n}[I \otimes I] = I \otimes I$.
Without loss of generality, $P_1 \neq I$.  
First consider $P_1 \neq P_2$.
There exists a $\tilde{c}$ such that $\tilde{c} P_1
\tilde{c}^\dagger = \sig{x}{1}$ and $\tilde{c} P_2 \tilde{c}^\dagger =
g=\sig{x}{2}$ or $\sig{z}{1}$ or $I$
depending on whether $[P_1,P_2] = 0$ or $\{P_1,P_2\} = 0$ or $P_2 =I$.
Then
\bea
	{\cal T}_{{\cal C}_n}[P_1 \otimes P_2] & = & 
	\frac{1}{|{\cal C}_n|} \sum_{c \in {\cal C}_n} 
	(c P_1 c^\dagger) \otimes (c P_2 c^{\dagger}) 
\nonumber
\\	& = & \frac{1}{|{\cal C}_n|} \sum_{c \in {\cal C}_n} 
	(c \sig{x}{1} c^\dagger) \otimes (c g c^{\dagger})\,.
\eea
For every $c$, there
is another $c'$ (see Eq.~(\ref{switch})) such that $c' \sig{x}{1}
c'^\dagger = - c \sig{x}{1} c^\dagger$ and $c' g c'^{\dagger} = c g
c^{\dagger}$, hence the sum vanishes.
Now consider $P = P_1 \otimes P_1$. 
\bea
	{\cal T}_{{\cal C}_n}[P_1 \otimes P_1] = 
	\frac{1}{|{\cal C}_n|} \sum_{c \in {\cal C}_n} 
	(c \sig{x}{1} c^\dagger) \otimes (c \sig{x}{1} c^\dagger)\,. 
\eea
Following the discussion on specifying Clifford group elements, $c
\sig{x}{1} c^\dagger$ ranges over all elements in $\tilde{\cal
P}_n-\{I\}$.  Moreover, each $c \sig{x}{1} c^\dagger$ occurs in the
sum the same number of times independent of $c$; this is the number of
valid combinations $a_2,\cdots,a_n$, $b_1,\cdots,b_n$ that complete
the image set, which is independent of $c$.  Thus we obtain
\be
	P \otimes P \stackrel{{\cal T}_{{\cal C}_n}}{\rightarrow} 
	{1 \over |\tilde{\cal P}_n| - 1} 
	\sum_{Q \in \tilde{\cal P}_n-\{I\}} Q \otimes Q\,.
\ee  
Comparing this with Eq.~(\ref{lasth}), we see that this is a linear
combination of $I$ and $H_n$.  
This proves condition (4), and thus ${\cal T}_{{\cal C}_n} = {\cal
T}_{U(2^n)}$.  Therefore ${\cal T}_{{\cal C}_n}[|{\bf 0}\>\<{\bf 0}|] =
\rho_0^{(n)}$.

\subsubsection{Selecting and implementing a random element in the 
Clifford group}

To implement the Clifford twirl efficiently, we need to select and
implement random elements in the Clifford group.
Our method is based on the circuit construction of any Clifford group element 
in Ref.~\cite{gotthesis} by Gottesman.  
The essence of his construction is as follows. 
We choose the following generating set for the Clifford group ${\cal
C}_n$:
\beq
	G=\{{\sc h}_i, {\sc cnot}_{jk}, {\sc p}_m, 
	{\sc p}^{\dagger}_n\}\label{clif}
\,, 
\eeq
where {\sc h}, {\sc cnot}, and {\sc p} respectively stand for the
1-qubit Hadamard transform $\frac{1}{\sqrt{2}}{1\,\,\,\,\,1\choose
1\,-1}$, the 2-qubit controlled-{\sc not}, and the 1-qubit phase gate
$1\,\,0\choose 0\,\,i$. Subscripts in Eq.~(\ref{clif}) denote the
qubit(s) being acted on.

Note that the generating set is self-inverse and that it has $n^2 + 2
n$ elements.  From Ref.~\cite{gotthesis}, a circuit with no more than
$3 n^2 + 7 n + O(1)$ gates from $G$ can be explicitly constructed for
each element in ${\cal C}_n$.

It remains to find a method to select any random element in ${\cal
C}_n$ with uniform probability.  This cannot be done simply by
randomizing the building blocks of the Gottesman construction.
Hence we use a random walk over ${\cal C}_n$ to generate a random
element, which can then be implemented by the Gottesman construction.
Even though $|{\cal C}_n|$ is of order $2^{O(n^2)}$, it can be proved
that our random walk converges in $O(n^8)$ steps to the uniform
distribution over the Clifford group elements.

The algorithm can be summarized as follows: 
\medskip
\begin{enumerate}
\item 
{\it Classical random walk.}~~Determine a random element using a
random walk algorithm: at every time step, with probability $1/2$, do
nothing, and with probability $1/2$ choose a random element in $G$.
Proceed for $O(n^8)$ steps.
Let the resulting element be $U$.
Classically compute the images $U \sig{x}{i} U^\dagger$ and $U
\sig{z}{i} U^\dagger$, which can be done efficiently following the
Knill-Gottesman theorem~\cite{gotthesis}.
\item 
{\it Quantum circuit.}~~Using the $2n$ images, build a quantum circuit to
implement $U$ using the Gottesman construction with $3 n^2 + 7 n + O(1)$
generators.
\end{enumerate}
\medskip
{\em Remarks}: We separate the algorithm into classical and quantum
parts because there are $O(n^8)$ classical steps but only $O(n^2)$
quantum gates. The `do nothing' step with probability 1/2 is based on
a technicality in the proof and may be skipped in an implementation,
so that a random generator is picked at every round.

\medskip

We prove that the Markov random walk `mixes' in $O(n^8)$ steps. The
proof~\cite{jer_sincl} relies on the facts that (1) any element of the
Clifford group can be reached from any other by applying $O(n^2)$
generators, i.e. the diameter $d$ of the (Cayley) graph of the group
is $O(n^2)$, and (2) the random walk uses a symmetric
(self-invertible) set of generators over a group. We use Corollary 1
in Ref.~\cite{dia_sal} to bound the second largest eigenvalue
$\lambda_2$ of this Markov chain as
\beq
\lambda_2 \leq 1-\eta/d^2\,,
\eeq
where $d$ is the diameter and $\eta$ is the probability of the least
likely generator, which is $\eta=\frac{1}{n^2+2n}$ in our case.
Therefore $\lambda_2 \leq 1-O(1/n^6)$. 
Then, using Lemma 2 in Ref.~\cite{dia_sal}, after $k$ steps of
iteration the distance of the obtained distribution $p^{(k)}$ from the
uniform distribution $u(c)=\frac{1}{|{\cal C}|}$ is bounded as
$||p^{(k)}-u||_1 \leq \sqrt{|{\cal C}|}(1-O(1/n^6))^k$. Here, we use
the $L_1$ norm between two distributions $p(c)$ and $q(c)$,
i.e. $||p-q||_1=\sum_{c \in {\cal C}}|p(c)-q(c)|$.  If we set
$k=O(n^8)$, this distance can be bounded by a small constant.

\subsection{Preparing $\rho_1^{(n)}$}

The state $\rho_1^{(n)}$ can be created using one singlet state.
This is obvious when $n=1$.  
For $n \geq 2$, $\rho_1^{(n)}$ can be created using the recurrence
relation Eq.~(\ref{recur2}) by the following recursive process:
\begin{itemize}
\item Flip a coin with bias $p_n$ for $0$, and bias
$1-p_n$ for $1$.
\item If the outcome is $0$, prepare $\rho_0^{(n-1)} \otimes
\rho_1^{(1)}$.  If the outcome is $1$, prepare 
$\rho_1^{(n-1)} \otimes \rho_0^{(1)}$.  
When $n-1 \geq 2$, $\rho_1^{(n-1)}$ is prepared recursively.  
Otherwise, $\rho_1^{(n-1)}$ is just the singlet.
\end{itemize}

Note that the procedure relies on the preparation of all possible  
$\rho_0^{(n)}$ without entanglement. 
Note also that the singlet $\rho_1^{(1)}$ is used exactly once in the
procedure. 


\section{Data hiding in separable states}
\label{sec:altsep}

We have considered hiding states with very little entanglement.  In this
section, we consider completely {\em separable} hiding states. 
Such ``separable hiding schemes'' are interesting for several reasons.
First, given a separable scheme to hide one bit, multiple bits
can be hidden bitwise; if the probability distributions of these bits are
independent, then the attainable information is additive (as will be 
proved in Section~\ref{sec:sepadd}).
Second, the hiding states can be prepared without entanglement.
Third, from a more fundamental perspective, such separable hiding
schemes exhibit the intriguing phenomenon of quantum nonlocality
without entanglement~\cite{qne} to the fullest extent.
We have good candidates for separable hiding states, but have not been
able to prove their security rigorously.
First we present a proof due to Wootters\footnote{Email correspondence
from W. K. Wootters, January 1999.} of the additivity of information
in a bitwise application of separable hiding schemes.

\subsection{Additivity of information when states are separable}
\label{sec:sepadd}

Suppose the bit $b=0,1$ can be hidden in the separable states
$\rho_{b=0,1}$, with bounded attainable mutual information, $I(B:Y) \leq
\delta$.  We call this the ``single-bit protocol''. 
We consider hiding two bits $b_1,b_2$ in the tensor-product state
$\rho_{b_1} \otimes \rho_{b_2}$.  Consider $I(B_1 B_2:Y_{12})$ where
$Y_{12}$ is the outcome of any measurement on $\rho_{b_1} \otimes
\rho_{b_2}$ which is LOCC between Alice and Bob but can be jointly on
$\rho_{b_1}$ and $\rho_{b_2}$.
By the chain rule of mutual information~\cite{cover&thomas:infoth}, we have
\beq 
	I(B_1 B_2:Y_{12}) = I(B_1:Y_{12}) + I(B_2:Y_{12}|B_1) 
\,.\label{adder}
\eeq 
We now show that, when $b_1$ and $b_2$ are independent (having a
product distribution $p(b_1)p(b_2)$), both $I(B_1:Y_{12})$ and
$I(B_2:Y_{12}|B_1)$ can be reinterpreted as the information obtained
about a bit hidden with the single-bit protocol and are bounded by
$\delta$. 
The term $I(B_1:Y_{12})$ measures how much about $b_1$ is learned from
$Y_{12}$, when $b_2$ is unknown.
This is also the information about $b_1$ learned from $\rho_{b_1}$ by
the following LOCC procedure: First, append an extra $\rho_{b_2}$,
chosen according to the probability distribution $p(b_2)$.  Then,
without using their knowledge of $b_2$, Alice and Bob measure $Y_{12}$
on $\rho_{b_1} \otimes \rho_{b_2}$.
Thus $I(B_1:Y_{12}) \leq \delta$.  It is crucial that $\rho_{b_2}$ is
separable and can thus be prepared by LOCC.
Similarly,
$I(B_2:Y_{12}|B_1)$ is the information about $b_2$ learned from
$\rho_{b_2}$ by appending a {\em known} extra $\rho_{b_1}$ (chosen
with probability $p(b_1)$) and measuring $Y_{12}$.  Hence
$I(B_2:Y_{12}|B_1) \leq \delta$.
More generally, $I(B_1 B_2 \cdots B_k:Y) \leq k \delta$ for any $Y$,
implying that the information is additive.

Note that this additive information bound for separable hiding states
is much stronger than that for entangled hiding states
(Section~\ref{sec:multiple}).  Note that there is an important
difference between hiding with separable states and the problem of
classical data transmission through a (noisy) quantum channel.  For
the latter it is known that the capacity is nonadditive, in the sense
that the receiver has to perform joint measurements on the data to
retrieve the full Holevo information \cite{holevo:cap}.  The
difference is that to achieve the Holevo information one encodes the
classical data, so that the prior distribution is not independent over
the different states.  In our information bound for bit hiding, the
different bits are assumed to have independent prior probabilities.

\subsection{An alternative hiding scheme}
\label{sec:tal}

In this section, we discuss some interesting properties of two
orthogonal separable states in ${\cal H}_2 \otimes {\cal H}_2$, and a
candidate separable hiding scheme built from them.  Consider the
following two bipartite states in ${\cal H}_2 \otimes {\cal H}_2$,
introduced in Section VII of~\cite{qne}:
\bea 
	\tau_0 	= \frac{1}{2} \lbm |+\>\<+| \otimes |0\>\<0| 
				  + |0\>\<0| \otimes |+\>\<+| \rbm \,,
\nonumber
\\
	\tau_1 	= \frac{1}{2} \lbm |-\>\<-| \otimes |-\>\<-| 
				  + |1\>\<1| \otimes |1\>\<1| \rbm \,,
\eea
where $|\pm\> = {1 \over \sqrt{2}} (|0\> \pm |1\>)$.  
These separable states are orthogonal, but the results of
Ref.~\cite{qne} suggest that they are not perfectly distinguishable by
LOCC.
We now strengthen this result using the general framework for LOCC
measurements described in Section~\ref{sec:povm}.

Our goal as before is to bound $p_{0|0} + p_{1|1}$ obtained by any
measurement with PPT POVM elements.  Following the discussion of
Section~\ref{sec:povm}, we restrict to PPT POVM elements $M_{0,1}$,
and use the symmetry of $\tau_{0,1}$ to simplify the possible form of
$M_{0,1}$.
It will suffice to consider POVM elements ${\cal T}^\dagger[M_i]$,
where ${\cal T}[\rho] = {1 \over 4} (\rho + {\sc s} \rho {\sc s} +
{\sc h}_2 \rho {\sc h}_2 + {\sc s} {\sc h}_2 \rho {\sc h}_2 {\sc s})$
with ${\sc h}_2 = {\sc h} \otimes {\sc h}$ being the bitwise Hadamard
transformation on both qubits, and {\sc s} being the swap operation on
the two qubits.  It is immediate that $\cal T$ is
self-adjoint, and that it fixes $\tau_{0,1}$.  Unlike in
Section~\ref{sec:povm}, $\cal T$ is not LOCC, but it does preserve the
PPT property of $M_i$, because ${\sc s}M_i{\sc s}$ is PPT (the swap
just relabels the input bits).  Moreover, $\forall M$,
${\cal T}[{\cal T}[M]] = {\cal T}[M]$; therefore, we can restrict to
POVM elements $M$ that are invariant under ${\cal T}$.  This symmetry
is most easily imposed in the Pauli decompositions of $M_{0,1}$:
\bea
	M_0 & = & a P_a + c P_c + e P_e + d P_d \,,
\nonumber
\\
	M_1 & = & (1-a) P_a - c P_c - e P_e - d P_d \,,  
\label{paulim1}
\eea
where 
\bea
	P_a & = & I \otimes I \,,
\nonumber
\\
	P_c & = & \sigma_z \otimes I + I \otimes \sigma_z 
 		+ \sigma_x \otimes I + I \otimes \sigma_x  \,,
\nonumber
\\
	P_d & = & \sigma_z \otimes \sigma_z + \sigma_x \otimes \sigma_x \,,
\nonumber
\\
	P_e & = & \sigma_z \otimes \sigma_x + \sigma_x \otimes \sigma_z \,,
\eea
are the only linearly independent invariants under ${\cal T}$.  
Note that $M_{0,1}$ are automatically invariant under partial
transpose.  Therefore, it only remains to impose
conditions on $a,c,d,e$ to make $0 \leq M_0 \leq I$:
\bea 
	0 \leq a - 2d \leq 1 \,,
\label{c2}
\\
	0 \leq a - 2e \leq 1 \,,
\label{c3}
\\
	0 \leq \alpha \pm \beta  \leq 1 
\,.
\label{c4}
\eea
In Eqs.~(\ref{c2})-(\ref{c4}), the bounded quantities are 
eigenvalues of $M_0$, with $\alpha = a+d+e$ and $\beta = \sqrt{8 c^2 +
(d-e)^2}$.

To find $p_{0|0}$ and $p_{1|1}$, we express $\tau_{0,1}$ in their
Pauli decompositions, using $|0\>\<0| = {1\over2}(I + \sigma_z)$,
$|1\>\<1| = {1\over2}(I - \sigma_z)$, and $|\pm\>\<\pm| = {1\over2}(I
\pm \sigma_x)$:
\bea
	\tau_0 = {1 \over 8} ( 2 P_a + P_c + P_e) \,,
\nonumber
\\
	\tau_1 = {1 \over 8} ( 2 P_a - P_c + P_d) \,.
\label{paulit1}
\eea 
Using Eqs.~(\ref{paulim1}) and (\ref{paulit1}), and the trace
orthonormality of the Pauli matrices (up to a multiplicative
constant), we find the conditional probabilities of interest:
\bea
	& p_{0|0} = \mbox{Tr}(M_0 \tau_0) & \!\!\!\! = a + 2c + e \,,
\nonumber
\\
	& p_{1|1} = \mbox{Tr}(M_1 \tau_1) & \!\!\!\! = (1-a) + 2c - d \,. 
\eea
\begin{figure}[f]
\begin{center}
\centerline{\mbox{\psfig{file=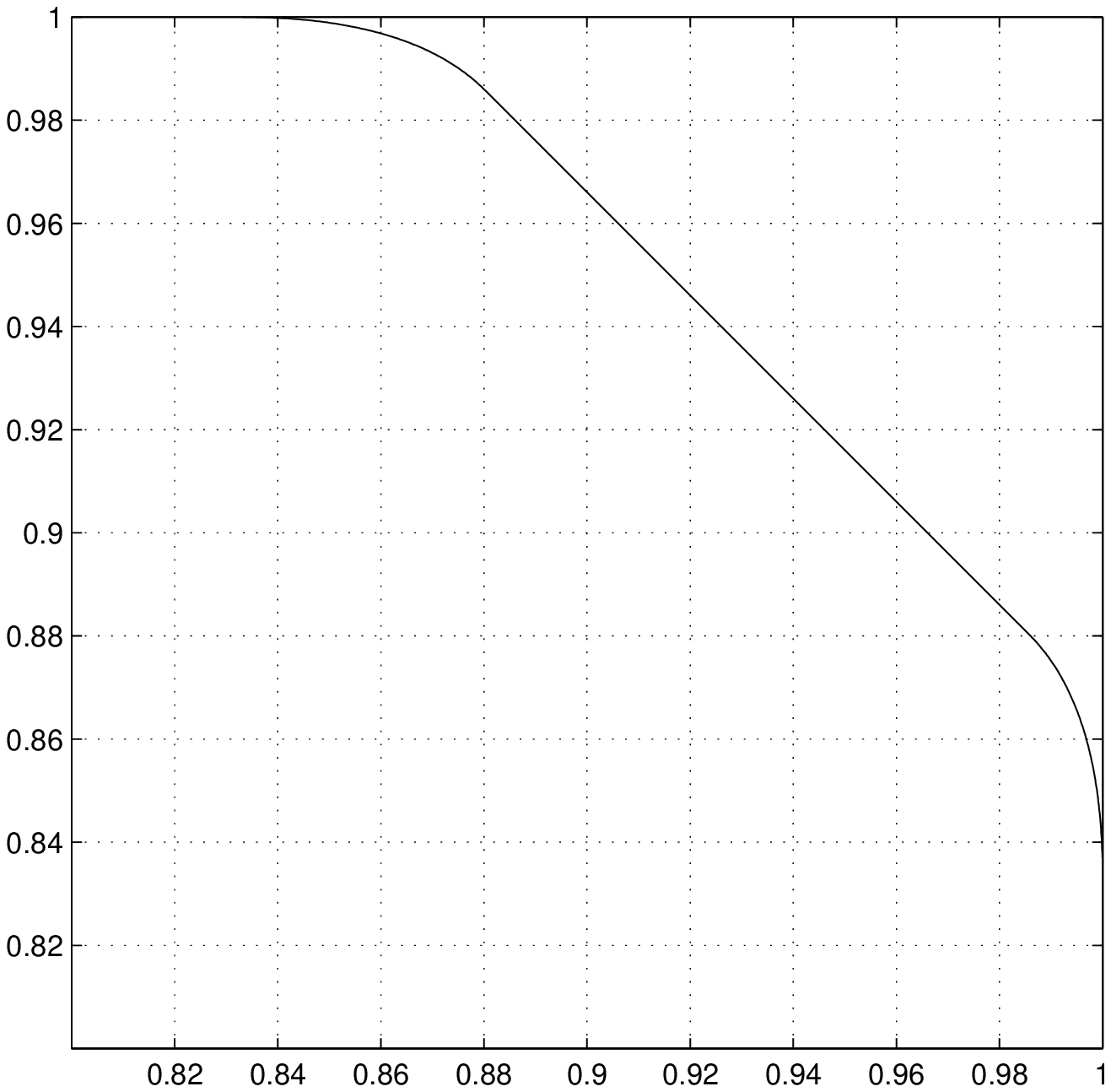,width=3in}}}
\vspace*{0.4cm}
\caption{$(p_{0|0}^{(1)}, p_{1|1}^{(1)})$ attainable by PPT-preserving 
measurements on $\tau_{0,1}$.}
\label{fig:talppt}
\end{center}
\end{figure}
The values of $(p_{0|0}, p_{1|1})$ permitted by the positivity
constraints Eqs.~(\ref{c2})-(\ref{c4}) are depicted in
Fig.~\ref{fig:talppt}; we leave the straightforward derivation of
Fig.~\ref{fig:talppt} to the interested reader.  Here we only use
Eq.~(\ref{c4}) to derive a simple bound on 
$\left| p_{0|0} + p_{1|1} - 1 \right|$
(the straight portion of the boundary).
We have: 
\bea 
\hspace*{-3ex} \left| p_{0|0} + p_{1|1} - 1 \right| \hspace*{-2ex} & = & 
\hspace*{-2ex} 4c + (e-d) \nonumber
\\
	\hspace*{-2ex} & = & \hspace*{-2ex} \sqrt{3} \, 
	|\beta| \left| \rule{0pt}{3.0ex} \right. \! 
	\sqrt{{2 \over 3}} {\sqrt{8} c \over |\beta|}
	 + {1 \over \sqrt{3}} {(e-d) \over |\beta|} 
	\! \left. \rule{0pt}{3.0ex} \right| \,.
\eea
Since $\beta = \sqrt{8 c^2 + (d-e)^2}$, the last factor in 
the last line is of the form $|\cos \theta \cos \phi + 
\sin \theta \sin \phi | = |\cos(\theta - \phi)| \leq 1$.  
Moreover, from Eq.~(\ref{c4}), $|\beta| \leq {1 \over 2}$.  
Hence, 
\be 
	\left|p_{0|0} + p_{1|1}-1 \right| \leq {\sqrt{3} \over 2}
\,.
\label{talbdd}
\ee
This upper bound for protocols with the P-H property is tight, in the
sense that there exists a simple LOCC procedure that achieves this
bound.  This procedure is described in Appendix~\ref{sec:talapp}.

Equation~(\ref{talbdd}) establishes that $\tau_{0,1}$ {\em cannot} be
perfectly distinguished by measurements with PPT POVM elements, since
the perfect measurement would give $p_{0|0} + p_{1|1} = 2$.
We can also put a lower bound on the amount of entanglement required 
to distinguish $\tau_0$ from $\tau_1$ perfectly. 
The combined support of $\tau_0$ and $\tau_1$ has full rank and
$\tau_0$ and $\tau_1$ are orthogonal; therefore, the perfect POVM
measurement has unique $M_{0,1}$.
The measurement, if used as in Fig.~\ref{fig:fig1}(b), can create the
states $M_0/{\rm Tr}(M_0)$ and $M_1/{\rm Tr}(M_1)$.  These both have
$\approx 0.55$ ebits of entanglement of formation, as calculated using
Ref.~\cite{woot}.  Therefore, $\tau_{0,1}$ take at least $0.55$ ebits
to distinguish perfectly.

\begin{figure}[f]
\begin{center}
\centerline{\mbox{\psfig{file=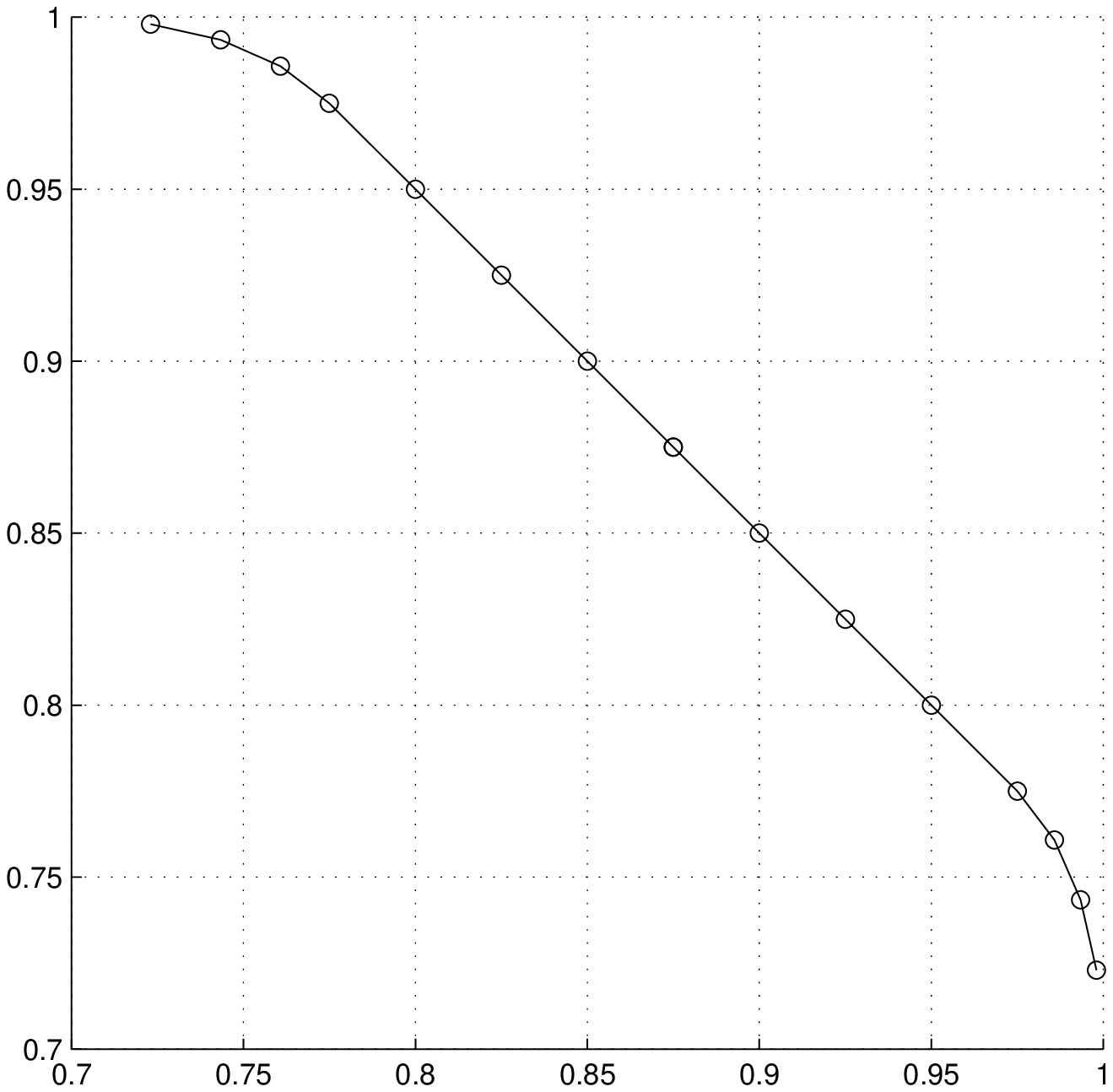,width=2.6in}}}
\vspace*{0.4cm}
\caption{$(p_{0|0}^{(2)}, p_{1|1}^{(2)})$ attainable by PPT-preserving 
measurements on $\tau_{0,1}$.}
\label{fig:tal2bdd}
\end{center}
\end{figure}

We conjecture that the parity of the number of $\tau_1$ in a tensor
product of $n$ $\tau_0$s and $\tau_1$s cannot be decoded better than
by measuring each tensor component and combining the results.
More precisely, we consider distinguishing the states 
\be
	\tau_b^{(n)} = {1 \over 2^{n-1}} 
	\sum_{b_1 \oplus b_2 \oplus \cdots \oplus b_n = b}
	\tau_{b_1} \otimes \tau_{b_2} \otimes \cdots \otimes \tau_{b_n}
\,. 
\ee
Here $\oplus$ is addition modulo two.
We follow the general method to find the allowed values of
$(p_{0|0}^{(n)},p_{1|1}^{(n)})$, and consider PPT $M_{0,1} \geq 0$
with the proper symmetries.
Since the eigenvalues of $M_0$ for $n \geq 2$ are not analytically
obtainable, we have performed a numerical maximization of $p_{1|1}^{(n)}$
with fixed $p_{0|0}^{(n)}$ for $n=2$ and $n=3$.
All numerical results presented have negligible numerical errors.
The allowed region for $(p_{0|0}^{(2)},p_{1|1}^{(2)})$ is given in
Fig.~\ref{fig:tal2bdd}.
The best value of $p_{0|0}^{(2)} + p_{1|1}^{(2)}$ is {\em precisely}
the one achieved by applying the LOCC measurement in
Appendix~\ref{sec:talapp} on each tensor component, and classically
combining the results to infer the parity.
The same has been confirmed for $n=3$: $p_{0|0}^{(3)} + p_{1|1}^{(3)}
\leq 1.64952$.
While we have no convincing arguments for the conjecture for general
$n$, the $n=2,3$ cases are unlikely to be degeneracies or
coincidences.

Suppose the conjecture is true.  Let ${1 \over 2} (p_{0|0}^{(1)} +
p_{1|1}^{(1)}) \leq p$ where $p = {1 \over 2} + {\sqrt{3} \over 4}$ 
denotes the best decoding probability for $n=1$.
Then, $\tau_b^{(n)}$ is distinguished correctly only if an even number
of components are decoded incorrectly, thus
\bea
	p_{0|0}^{(n)} + p_{1|1}^{(n)} 
	& \leq & 2 \sum_{k~{\rm even}} p^{n-k} (1-p)^k {n \choose k}
\nonumber
\\
	& = & (p + (1-p))^n + (p - (1-p))^n 
\nonumber
\\
	& = & 1 + (2 p-1)^n 
\nonumber
\\
	& = & 1 + \left(\rule{0pt}{2.4ex} \right. \!
		{ \sqrt{3} \over 2 } \! \left. \rule{0pt}{2.4ex} \right)^n
\,.
\label{maybound}
\eea
The upper bound on $p_{0|0}^{(n)} + p_{1|1}^{(n)} - 1$ implies  
$\left| \rule{0pt}{2.1ex} \right. \!\!
p_{0|0}^{(n)} + p_{1|1}^{(n)} - 1 
\!\! \left. \rule{0pt}{2.1ex} \right| \leq 
\left(  \rule{0pt}{2.1ex} \right. \!\!  { \sqrt{3} \over 2 } 
\!\! \left. \rule{0pt}{2.1ex} \right)^n$.  
If the conjecture is true, this bound vanishes exponentially with the
number of qubits used, and $\tau_b^{(n)}$ can be used to hide a single
bit in a way similar to the Bell mixtures $\rho_b^{(n)}$.  Comparing
Eqs.~(\ref{maybound}) and (\ref{cleanbound}), the separable scheme
takes ${1\over1-\log_2\sqrt{3}}\approx 4.8$ times as many qubits as in
the Bell-state protocol to achieve the same level of security.

\section{Conditionally Secure Quantum Bit Commitment}
\label{sec:qbc}

Bit commitment~\cite{kilian} is a cryptographic protocol with two parties,
Alice and Bob.  It has two stages, the commit and the open phases.
The goal is to enable Alice to commit to a bit that can neither be
learned by Bob before the open phase nor be changed by Alice after 
the commit phase. 
Bit commitment is a primitive for many other protocols, 
such as coin tossing, cf. Ref.~\cite{kilian}. 
However, the security of classical bit commitment schemes relies on
unproved assumptions on computational complexity, while
unconditionally secure quantum bit commitment has been proved to be
impossible~\cite{Mayers97bc,Lo97bc}.

In this section, we use the main idea of the bit hiding scheme in
Section~\ref{sec:bithid1} to construct a conditionally secure bit
commitment scheme in the following setting.
The commitment is shared between two recipients (Bob-1 and Bob-2) who
do not share entanglement and can perform only LOCC.
Thus the present discussion contrasts with previous works in that (1) it
is not precisely a two-party protocol and (2) the security is {\em
conditioned} on restricting the two Bobs to LOCC operations only.

Let $n,r$ be security parameters.  The scheme is secure in the sense
that the dishonest Bobs learn at most $2^{-(n-1)}$ bits on the committed bit
before the open phase, and a dishonest Alice can change her commitment
without being caught with probability $2^{-r}$.
The scheme is as follows:
\vspace*{1ex}
\begin{itemize}
\item Commit phase: To commit to $b = 0$ ($b=1$), Alice picks a random
$|w_{\bf k}\>$ with\footnote{Recall that $|w_{\bf k}\>$ denotes a
tensor product of $n$ Bell states specified by the $2n$-bit string
{\bf k} according to the scheme in Eq.~(\ref{iden}).}
an even (odd) number of singlets $|w_{11}\>$.
Alice sends one qubit of each Bell pair to each Bob.
\item Open phase: Alice sends $n+r$ singlets to Bob-1 and Bob-2, who
apply a random hashing test~\cite{bdsw,lo&chau:security} and use $r$
of the singlets to check if the received states are indeed singlets.
Alice is declared cheating if the test fails.
Otherwise, the remaining $n$ singlets are used to teleport Bob-1's
qubits to Bob-2 who measures {\bf k} to find $b$.
\end{itemize}

\medskip
\begin{proof} [Security] If Alice is honest, the security of the bit
hiding scheme implies that the dishonest Bobs can learn at most
$O(2^{-n})$ bits of information before the open phase.
If Bob is honest, the most general strategy for a dishonest Alice is
to prepare a {\em pure} state $|\psi\>$ with five parts, $P_i$ for
$i=1,\cdots,5$.  In the commit phase, she gives $P_1$ and $P_2$ to
Bob-1 and Bob-2.  In the open phase, she applies some quantum
operation $\cal E$ on $P_3,P_4,P_5$.  Then, she sends $P_3$ and $P_4$
to Bob-1 and Bob-2.
If $P_3,P_4$ are indeed $n+r$ singlets, the test is passed and
$\cal E$ does not change the state of $P_1$ and $P_2$, and Bob-2
indeed obtains $P_1,P_2$ in the same state as in the commit phase
after teleportation.  ($P_1,P_2$ can only be changed during the
teleportation steps when interacting with $P_3,P_4$ if they are not 
singlets.)
In this case, the distribution of {\bf k} and therefore the committed
bit $b$ in both phases are the same -- Alice cannot change or delay
her commitment.
In case $P_3$ and $P_4$ are not singlet states the analysis of the
failure probability of the random hashing method follows the quantum
key distribution security proof by Lo and Chau
\cite{lo&chau:security}. If $P_3,P_4$ are in some state orthogonal to
$n+r$ singlets, the test is passed with probability $2^{-r}$ only.
In general, let $\alpha$ be the fidelity of $P_3,P_4$ with respect to
$n+r$ singlets.
The probability to change the commitment without being caught is $\leq
2^{-r} (1-\alpha)$ which is less than $2^{-r}$. 

\end{proof}
Note that our scheme does not force Alice to commit.  
For example, she can send an equal superposition of the $b=0,1$
states, but she cannot control the outcome at the opening phase.  
However, this is inherent to schemes which conceal the commitment 
from Bob.
Classically, it is as if Alice sent an empty locked box with no bit
written inside, or a device in which Bob's opening the box triggered a
fair coin toss to determine the bit.

We note that the above scheme has many equivalent variations.  For
example, Alice may instead prepare a superposition $\sum_{{\bf k} \in
E_n/O_n} \ket{\bf k} \otimes \ket{w_{\bf k}}$ and send the second part
to the Bobs, which provides a way of sending the density matrices
$\rho_b^{(n)}$.  Bob can request Alice to announce {\bf k} in the open
phase, and declare her cheating if he finds a different {\bf k}.
However, these schemes are exactly equivalent to the proposed one. 

\section{Discussion}
\label{sec:disc}

The quantum data hiding protocol using Bell mixtures can be
demonstrated with existing quantum optics techniques.
The hider can prepare any one of the four polarization Bell states, 
$\frac{1}{\sqrt{2}}(\ket{\updownarrow\,,
\updownarrow}\pm\ket{\leftrightarrow\,,\leftrightarrow})$,
$\frac{1}{\sqrt{2}}(\ket{\updownarrow\,,
\leftrightarrow}\pm\ket{\updownarrow\,,\leftrightarrow})$.
Any of these can be prepared using
downconversion~\cite{mattleetal}, followed by an appropriate single 
photon operation. 
The photons can be sent through two different beam paths to Alice and Bob.
Then Alice and Bob might attempt to unlock the secret by LOCC
operations as in Section~\ref{tightb}; this requires high efficiency
single-photon detection.\footnote{
The quantum efficiency of detectors strongly depends on the
wavelength.  At $543$ nm, quantum efficiencies as high as $95$\% have
been observed (E.~Waks {\it et al.}, unpublished).  At $694$ nm,
quantum efficiencies as high as $\approx 88$\% were reported,
see Ref.~\protect\cite{Takeuchi99}.}
Alternatively, the secret can be completely unlocked if a quantum
channel is opened up for Alice to send her photons to Bob, who
performs an incomplete measurement on each pair to distinguish the
singlet $\frac{1}{\sqrt{2}}(\ket{\updownarrow\,,
\leftrightarrow}-\ket{\updownarrow\,,\leftrightarrow})$ from the
other three Bell states.  Such an incomplete measurement has been
performed in the lab~\cite{mattleetal}; a full Bell measurement is not
necessary and is in fact not technologically feasible in current
experiments.

Our alternative low-entanglement preparation scheme would require more
sophisticated quantum-optics technologies than presently exist but is
interesting to consider.  As we showed in Section~\ref{sec:prepstate},
quantum operations in the Clifford group are required.  The one-qubit
gates are obtainable by linear optics, but the {\sc cnot} gate cannot
be implemented perfectly by using linear optical elements.  However,
recent work by Knill {\em et al.}~\cite{klm} shows that a {\sc cnot}
gate can be implemented near-deterministically in linear optics when
single-photon sources are available.  So, this preparation scheme may
be practicable in the near future as well.

We have proved for the Bell mixtures $\rho_{0,1}^{(n)}$ that the
secret is hidden against LOCC measurements; but the secret can be
perfectly unlocked by LOCC operations if Alice and Bob initially share
$n$ ebits of entanglement (the LOCC procedure is just quantum
teleportation followed by a Bell measurement by Bob).
What is the smallest amount of entanglement $E_{min}$ required for
perfect unlocking?  This quantity would give another interesting
measure of the strength of hiding.  We do not know $E_{min}$ exactly,
although we can give a lower bound for it.
The bound is obtained by considering the procedure of
Fig.~\ref{fig:fig1}(b) as a state-preparation protocol.  Here we take
the measurement {\bf M} to be the nonlocal one that {\em exactly}
distinguishes the hiding states.
This {\bf M} is unique because $\rho_{0,1}^{(n)}$ have full combined
support and are orthogonal.
Considering {\bf M} to be an LOCC measurement performed with the
assistance of $E_{min}$ ebits of prior entanglement, the average
output entanglement of formation of Fig.~\ref{fig:fig1}(b) provides a
lower bound for $E_{min}$, since the LOCC procedure cannot increase
the average entanglement.

This entanglement of formation is straightforward to compute: When the
input is two halves of two maximally entangled states as shown, the
input density matrix, which is proportional to the identity, can be
expressed as:
\be
        {I \over 4^n} = 
        {|E_n| \over 4^n} \rho_0^{(n)} + 
        {|O_n| \over 4^n} \rho_1^{(n)} 
\,.
\ee
The output state is therefore $\rho_{0,1}^{(n)}$ with probabilities
${|E_n| \over 4^n}$, ${|O_n| \over 4^n}$.  Since the entanglement of
formation of $\rho_{1}^{(n)}$ is 1 ebit, the average output
entanglement of formation is ${|O_n| \over 4^n} = {1 \over 2} (1 -
2^{-n})$ ebits.  Thus, $E_{min}\geq {1 \over 2} (1 - 2^{-n})$.  This
is still far from the upper bound of $n$ ebits; considering more
general input states in Fig.~\ref{fig:fig1}(b) could improve our lower
bound.

Our technique of putting bounds on the capabilities of LOCC operations
using the Peres-Horodecki criterion is an application of the work of
Rains~\cite{rainsimp,rainsdist}.  He introduced a class of quantum
operations that contains the LOCC class, which he called the {\em
p.p.t.~superoperators}~\cite{rainsimp,rainsdist}; $\cal S$ is in the
p.p.t.~class iff $({\bf 1}\otimes T)\circ{\cal S}\circ({\bf 1}\otimes
T)$ is a completely positive map.  It has been
proved\footnote{E. M. Rains, private communication, February 2001.}
that the p.p.t.~class is identical to the class of quantum operations
that have the Peres-Horodecki property, which we introduce in
Section~\ref{sec:povm}.

This p.p.t.~class also extends in a very precise way to the POVMs, in
the following sense: the condition that all $M_i$ be PPT is both a
necessary and sufficient condition for a POVM to have the P-H property
(which is equivalent to saying that the POVM, viewed as a quantum
operation from ${\cal H}_d\otimes {\cal H}_d$ to a one-dimensional
space ${\cal H}_1\otimes {\cal H}_1$, is p.p.t.).  The
necessary part is proved in Section~\ref{sec:povm}; sufficiency can be
derived by considering an arbitrary input $\rho$ to the POVM {\bf M}
including ancillas (as in Fig.~\ref{fig:fig1}(b), but with general
input).  The output density matrix for outcome $i$ is proportional
to $\rho_i\propto{\rm Tr}_p (M_i\rho)$, where the partial trace is over
the input Hilbert space to {\bf M}.  The partial transpose of this
operator can be written $({\bf 1}\otimes T)[\rho_i]\propto{\rm Tr}_p
[({\bf 1}\otimes T)[M_i]({\bf 1}\otimes T)[\rho]]$; the proof follows
straightforwardly from this formula.

It should be noted that the symmetrizing operation ${\cal T}^\dagger$
introduced in Section~\ref{sec:povm} is {\em not} restricted to the
p.p.t.~class.  
For example, ${\cal T}^\dagger$ in Section~\ref{sec:tal} is not in the
p.p.t.~class.
${\cal T}^\dagger$ only needs to be unital and has the property that
${\cal T}^\dagger[M]$ is PPT if $M$ is PPT.  The latter resembles,  
but is very different from the P-H property, which also requires $({\bf 1}
\otimes {\cal T}^\dagger)[M]$ to be PPT if $M$ is PPT.
The difference is reminiscent of the distinction between positive and
completely positive maps.

We believe that our quantum data hiding scheme using Bell states can
be extended to multiple parties. Consider for example the extension to
three parties Alice, Bob and Charlie, who are restricted to carrying
out LOCC operations amongst each other. Candidates for the hiding
states are the 3-party extensions of the Werner states, studied by
Eggeling and Werner~\cite{wern_egg}.

\section{Acknowledgments}


We thank Charles Bennett for the suggestion to consider data hiding
with the separable $\tau$ states and discussions on hiding with 
Bell states and quantum bit commitment.
We thank Bill Wootters for discussions and insights about the additivity of
information in the case of hiding in separable states.
We thank Vittorio Castelli for giving the proof relating error
probability and mutual information which we use in bounding the
attainable information about multiple bits.
We thank Don Coppersmith, Greg Sorkin and Dorit Aharonov for
discussion and for pointing out the relevant literature on random
walks on Cayley graphs with a small diameter.
We thank John Smolin for discussion and for writing the initial
program for the numerical work described in Section~\ref{sec:altsep}.
We acknowledge support from the National Security Agency and the
Advanced Research and Development Activity through the Army Research
Office contract number DAAG55-98-C-0041.

\appendices

\section{Bounding the obtainable mutual information}
\label{sec:mutualinfo}

Let $X$ be a binary random variable.  Define $Z_s$ to be the set of
all binary random variables that satisfy the following bound 
on the conditional probabilities: 
\begin{equation}
	-\delta \leq p(Z=0|X=0) + p(Z=1|X=1) - 1 \leq \delta
\,,
\label{constr}
\end{equation}
as in Eq.~(\ref{finalb}).  Define $Y_s$ as the set of all random
variables, with any number of outcomes, for which all ``decoding 
processes'' ${\cal D}$ produce binary random variables in the
set $Z_s$.  This situation is summarized as 
\beq
	X \stackrel{{\cal M}}{\rightarrow} Y 
			\stackrel{{\cal D}}{\rightarrow} Z\,.
\eeq
Note that $p(Z|Y)$ prescribed by $\cal D$ may depend on $p(Y|X)$.  
In this setting, we have the following bound:

\medskip
\begin{theo} $\forall Y \in Y_s,\ \ I(X:Y) \leq \delta H(X)$.
\label{intmutual}
\end{theo}
\medskip
\begin{proof} We first derive bounds on the conditional probability
distribution of any $Y \in Y_s$ given $X$.  We consider 
$p(Z\!=\! k|X\!=\! l) = \sum_j p_{{\cal D}}(Z\!=\! k|Y\!=\! j)
p(Y\!=\! j|X\!=\! l)$
implied by any $Y$ and $\cal D$. 
When we maximize $I(X:Y)$ over $Y$ with fixed ${\cal D}$, we obtain an
overestimate of $I(X:Y)$ because we can maximize $Y$ over a superset of
$Y_s$. 
Hence we can focus on a particular decoding process ${\cal D}$ defined as: 
\bea
	p_{\cal D}(Z\!=\! 0|Y\!=\! j)=1 & {\rm if} & j \in I_0 \cup I_- 
\,,
\nonumber
\\ 
	p_{\cal D}(Z\!=\! 1|Y\!=\! j)=1 & {\rm if} & j \in I_+
\,,
\eea
where
\bea
 j \in I_+ \Leftrightarrow p(Y\!=\! j|X\!=\! 1) > p(Y\!=\! j|X\!=\! 0)\,,
\nonumber
\\
 j \in I_0 \Leftrightarrow p(Y\!=\! j|X\!=\! 1) = p(Y\!=\! j|X\!=\! 0)\,,
\nonumber
\\
 j \in I_- \Leftrightarrow p(Y\!=\! j|X\!=\! 1) < p(Y\!=\! j|X\!=\! 0)\,.
\eea
Given these we can compute
\bea
	\lefteqn{p(Z\!=\! 0|X\!=\! 0)+p(Z\!=\! 1|X\!=\! 1)} \hspace{0cm} 
\nonumber
\\
	& = & \!\!\! \!\!
		\sum_{j \in I_0 \cup I_-} p(Y\!=\! j|X\!=\! 0)\! 
		+\! \sum_{j \in I_+} p(Y\!=\! j|X\!=\! 1) \hspace*{12ex} 
\nonumber
\\
	& = & \sum_j \max ( p(Y\!=\! j|X\!=\! 0),p(Y\!=\! j|X\!=\! 1) ) 
		\hspace*{12ex}
\nonumber
\\
	& = & \!\! 1\!+\! \sum_{j\in I_+}p(Y\!=\!j|X\!=\!1)-p(Y\!=\!j|X\!=\!0)
\,.
\label{deltabdd}
\eea
Substituting Eq.~(\ref{deltabdd}) into Eq.~(\ref{constr}), the first
inequality becomes trivial and the second inequality gives a necessary
condition on the set $Y_s$:
\\[1.5ex]
$~~\forall Y \in Y_s$, 
\begin{eqnarray}
	\sum_{j\in I_+} p(Y\!=\! j|X\!=\! 1)-p(Y\!=\! j|X\!=\! 0) \leq \delta
\,.
\label{nec1}
\end{eqnarray}

We now use Eq.~(\ref{nec1}) to bound $I(X:Y)$.  We introduce the
notations $p_{j0}=p(Y\!=\! j|X\!=\! 0)$ and $p_{j1}=p(Y\!=\! j|X\!=\! 1)$ 
for the conditional probabilities, and $p(X\!=\! 0)=x_0$ and
$p(X\!=\! 1)=x_1=1-x_0$ for the fixed prior probabilities for $X$.
We can then write the mutual information as
\bea
	& I(X:Y) = \sum_{j \in I_+ \cup I_-} f(p_{j0},p_{j1}) & 
\\
	\hspace*{-8ex} \mbox{where} & &  
\nonumber
\\
	& f(p_{j0},p_{j1}) = \sum_{\alpha=0}^1 x_\alpha p_{j\alpha}
	\log{p_{j\alpha} \over x_0 p_{j0} + x_1 p_{j1}}  & 
\label{bigmut}
\eea
represents the information on $X$ obtained from the outcome $Y\!=\!j$, 
weighted by $p(Y\!=\!j)$.  
Note also the outcomes in $I_0$ do not contribute to the
mutual information.  We will maximize Eq.~(\ref{bigmut}) subject
to the constraints 
\be
	\sum_j p_{j0} = 1 \,, ~~~ \sum_j p_{j1} = 1
\,,
\label{n1}
\ee
and
\be
	\sum_{j \in I_+} p_{j1} - p_{j0} \leq \delta
\,.
\label{kap2}
\ee

The maximization of $I(X:Y)$ is made tractable by noting that any
optimal $Y$ can be replaced by another $Y''$ with $|I''_+|, |I''_0|,
|I''_-| \leq 1$, and satisfying the same constraints 
Eqs.~(\ref{n1}) and (\ref{kap2}).
We prove this using the fact that $f(p_0,p_1)=f({\vec p})$ is
linear ($f(c{\vec p})=cf({\vec p})$) and convex, giving
\bea
	f( c_0 \, {\vec p} + c_1 \, {\vec q}) 
	& \leq & c_0 f({\vec p}) + c_1 f({\vec q})\nonumber
\\
	& = & f(c_0 \, {\vec p}) + f(c_1 \, {\vec q})
\,.
\label{cnvx2}
\eea 
Absorbing the nonnegative factors $c_{0,1}$ into the probability
vectors, this becomes simply \beq f({\vec p}+{\vec q})\leq f({\vec
p})+f({\vec q})\,.\label{cnvx} \eeq
The convexity of $f$ is proved by showing that the Hessian matrix
$\partial^2 f/\partial p_{\alpha}\partial p_{\beta}$ is positive
semidefinite (it is straightforward to show that its eigenvalues are
$0$ and $x_0 x_1 (p_{0}^2 + p_{1}^2) / (p_0 p_1 \sum_\alpha x_\alpha
p_{\alpha})$).

Given any $Y$, we first construct an intermediate $Y'$ as follows: 
For each outcome $j \in Y$ with unequal conditional probabilities
$(p_{j0},p_{j1})$, introduce two outcomes in $Y'$ with 
conditional probabilities: 
\bea
(0,p_{j1}-p_{j0}) ~\mbox{and}~ (p_{j0},p_{j0}) 
				~~~\mbox{if}~~~ j \in I_+\,\,
\nonumber
\\
(p_{j0}-p_{j1},0) ~\mbox{and}~ (p_{j1},p_{j1}) 
				~~~\mbox{if}~~~ j \in I_- \,. 
\nonumber
\eea
All outcomes in $I_0$ occur in $Y'$ unchanged.  
Note that the number of outcomes are such that 
$|I'_\pm| = |I_\pm|$ and $|I'_0| = |I_+| + |I_-| + |I_0|$.    
The constraints Eqs.~(\ref{n1}) and (\ref{kap2}) are satisfied by $Y'$
because the quantities involved are conserved by construction.
$I(X:Y') \geq I(X:Y)$ by applying Eq.~(\ref{cnvx}) to each
replacement.
Finally, as all $I'_+$ outcomes have $p_{j0}=0$, and all $I'_-$
outcomes have $p_{j1}=0$, we introduce the desired random variable
$Y''$ with just three outcomes, with conditional probabilities 
$(0,\sum_{j \in I'_+} p_{j1})$, $(\sum_{j \in I'_-} p_{j0},0)$,
and $(\sum_{j \in I'_0} p_{j0}, \sum_{j \in I'_0} p_{j0})$.  
$Y''$ still satisfies Eqs.~(\ref{n1}) and (\ref{kap2}), and the linearity
of $f$ implies $I(X:Y'')=I(X:Y')$.  

So, we have established that there exists an optimal $Y$ with
conditional probabilities $(p_{10},0)$, $(0,p_{21})$, and
$(p_{30},p_{30})$; these may be interpreted as the ``certainly 0'',
``certainly 1'', and ``don't know'' outcomes.  It is now trivial to
show that the best choice of these parameters consistent with the
constraints is given by $p_{10} = p_{21} = \delta$, $p_{30}=1-\delta$.
These parameters lead to the mutual information $I(X:Y)=\delta H(X)$,
proving the theorem.
\end{proof}

\section{Proof of Theorem~\ref{theogenlower}}
\label{prove}

\begin{proof}
We write the density matrices $\rho_{0,1}$ of the hiding states in the
Pauli decomposition
\be
	\rho_b = {1 \over 4^n} \sum_{\bf k} a_{b {\bf k}} \sigma_{\bf k}\,,
\ee
where the sum is over  all $16^n$ possible $4n$-bit strings ${\bf k}$,
and $\sigma_{\bf k}$ is identified with a tensor product of $2n$ Pauli
matrices as defined in Section~\ref{sec:bithid1}.
We restrict our choice of LOCC measurements to those that measure the
eigenvalues of a particular optimal $\sigma_{\bf s}$, which can be
$+1$ or $-1$.
This measurement is in the LOCC class because it is the product of the
eigenvalues of all the Pauli matrix components, which can be measured
locally and communicated classically to obtain the final result.
If we associate the outcomes $+1$ and $-1$ with $\rho_0$ and $\rho_1$
respectively, the POVM elements
are $M_{{\bf s}0} = {1 \over 2} (I+\sigma_{\bf s})$ and $M_{{\bf s}1}
= {1 \over 2} (I-\sigma_{\bf s})$.
Using the fact 
\be
	a_{b{\bf s}} =\mbox{Tr} \sigma_{\bf s} \rho_b\,,
\ee
we can calculate the conditional probabilities of interest:
\bea
	p(+1 | ~{\bf s}, \rho_b) = \mbox{Tr}(M_{{\bf s}0} \rho_b) = 
	{1 \over 2} (1 + a_{b {\bf s}})\,,
\nonumber
\\
	p(-1 | ~{\bf s}, \rho_b) = \mbox{Tr}(M_{{\bf s}1} \rho_b) = 
	{1 \over 2} (1 - a_{b {\bf s}})\,.
\label{def2}
\eea
In this notation, $p_{0|0} + p_{1|1} - 1$ is given by 
\bea
	p(+1 | ~{\bf s},\rho_0) + p(-1 | ~{\bf s},\rho_1) - 1 
	= {1 \over 2} (a_{0{\bf s}} - a_{1{\bf s}})\,.\label{firstgo}
\eea
If the right-hand side of Eq. (\ref{firstgo}) is negative, then we can always
do better by inverting the assignment of outcomes, flipping the sign of this
factor.  So, we can always achieve
\bea
	p(+1 | ~{\bf s},\rho_0) + p(-1 | ~{\bf s},\rho_1) - 1 
	= {1 \over 2} |a_{0{\bf s}} - a_{1{\bf s}}|\,.
\eea
Our goal is to establish a lower bound on this quantity due to
the orthogonality of $\rho_0$ and $\rho_1$.
Thus we consider 
\bea
	\min_{\rho_{0,1}:\rho_0 \perp \rho_1} \max_{\bf s}  
		(p_{0|0} + p_{1|1} - 1) 
\nonumber
\\      = 	
	\min_{\rho_{0,1}:\rho_0 \perp \rho_1} \max_{\bf s}  
		{1 \over 2} |a_{0{\bf s}} - a_{1{\bf s}}|\,.
\label{minmax}
\eea
Let $\rho_{0,1}$ be fixed, and ${\bf s}^*$, which depends on
$\rho_{0,1}$, be the corresponding ${\bf s}$ which maximizes ${1 \over
2} |a_{0{\bf s}} - a_{1{\bf s}}|$.  Let
\be 
	q_0 = \max(a_{0{\bf s}^*},a_{1{\bf s}^*}) \,,
	\hspace*{2ex}
	q_1 = \min(a_{0{\bf s}^*},a_{1{\bf s}^*}) \,.
\label{def3}
\ee
We can rephrase the optimization in Eq.~(\ref{minmax}) as a
minimization over $a_{b{\bf s}}$ for ${\bf s} \neq {\bf 0}$ ($a_{b{\bf 0}} =
1$ is fixed by the normalization of $\rho_{0,1}$), subject to the
following constraints:
\begin{enumerate}
\item Optimality of ${\bf s}^*$: 
\be
	q_0 - q_1 = \kappa_{\bf s} |a_{0{\bf s}} - a_{1{\bf s}}|
	\mbox{ where } \kappa_{\bf s} \geq 1\,.
\label{constraint1}
\ee
\item Orthogonality of $\rho_0$ and $\rho_1$, implying that  
	$\sum_{{\bf s} \neq {\bf 0}} a_{0{\bf s}} a_{1{\bf s}} = -1$, or  
\be
	\sum_{{\bf s} \neq {\bf 0},{\bf s}^*} 
	a_{0{\bf s}} a_{1{\bf s}} + q_0 q_1 = -1 \,.
\label{constraint2}
\ee
\end{enumerate}
Note that we do not impose the positivity of $\rho_{0,1}$, and 
obtain a valid, though possibly loose bound.
The above constraints are imposed by introducing the Lagrange multipliers 
$\lambda$ and $\lambda_{\bf s}$ for ${\bf s} \neq {\bf 0},{\bf s}^*$, 
transforming the problem to the unconstrained minimization:  
\bea
	\min \left( \rule{0pt}{3.0ex} \right.\!\!
	q_0 - q_1 
        - \sum_{{\bf s} \neq {\bf 0}, {\bf s}^*} \lambda_{\bf s} 
	\left[  \rule{0pt}{2.1ex}
	(q_0-q_1)-\kappa_{\bf s} |a_{0{\bf s}}-a_{1{\bf s}}| \right]   
\nonumber
\\
	- \lambda \left[ \rule{0pt}{3.0ex} \right. \! 
	\sum_{{\bf s} \neq {\bf 0}, {\bf s}^*}  
	a_{0{\bf s}} a_{1{\bf s}} + q_0 q_1 
	\! \left. \rule{0pt}{3.0ex} \right] 
	\!\! \left. \rule{0pt}{3.0ex} \right) 
\,.
\label{bgmax}
\eea
We can fix $q_0$ and minimize over $q_1$ and $a_{b{\bf s}}$ for ${\bf
s} \neq {\bf 0}, {\bf s}^*$.  If $a_{0{\bf s}}\neq a_{1{\bf s}}$
whenever ${\bf s}\neq {\bf 0}$ (the other case will be discussed
later) this function is analytic and we obtain the minimum by setting
the derivatives of Eq.~(\ref{bgmax}) with respect to the independent
variables $q_1$, $a_{0{\bf s}}$, and $a_{1{\bf s}}$ to zero:
\bea
	& 1 + \lambda q_0 - \sum_{{\bf s} \neq {\bf 0},{\bf s}^*} 
	\lambda_{\bf s} = 0\,, &
\label{first}
\\
	& r_{\bf s}\kappa_{\bf s} \lambda_{\bf s} + \lambda a_{1{\bf s}} = 0 
	& ~\forall~{\bf s} \neq {\bf 0},{\bf s}^*\,, 
\label{penult}
\\
	& -r_{\bf s}\kappa_{\bf s} \lambda_{\bf s} + \lambda a_{0{\bf s}} = 0 
	& \forall~{\bf s} \neq {\bf 0},{\bf s}^*\,.  
\label{last}
\eea
Here $r_{\bf s}=+1$ if $a_{0{\bf s}}>a_{1{\bf s}}$ and $r_{\bf
s}=-1$ if $a_{0{\bf s}}<a_{1{\bf s}}$.  We need to solve
Eqs.~(\ref{constraint1}), (\ref{constraint2}), (\ref{first}), and
(\ref{last}) for $q_0-q_1$.  First of all, we eliminate the $a_{b{\bf
s}}$ by substituting Eqs.~(\ref{penult}) and (\ref{last}) into the
constraints Eqs.~(\ref{constraint1}) and (\ref{constraint2}):
\bea
	& \lambda (q_0-q_1) = 2 \kappa_{\bf s}^2 \lambda_{\bf s}\,, & 
\label{resum}
\\
	& -\sum_{{\bf s} \neq {\bf 0},{\bf s}^*} 
	\kappa_{\bf s}^2 \lambda_{\bf s}^2 + \lambda^2 q_0 q_1 = 
-\lambda^2\,. & 
\label{aone}
\eea
We can obtain two other equations from Eq.~(\ref{resum}):
\bea
	\lambda (q_0-q_1) \sum_{{\bf s} \neq {\bf 0},{\bf s}^*} 
	\lambda_{\bf s}
	& = & 2 \sum_{{\bf s} \neq {\bf 0},{\bf s}^*} 
	\kappa_{\bf s}^2 \lambda_{\bf s}^2\,,
\nonumber
\\
	\lambda (q_0-q_1) \sum_{{\bf s} \neq {\bf 0},{\bf s}^*}
	{1 \over \kappa_{\bf s}^2} 
	& = & 2 \sum_{{\bf s} \neq {\bf 0},{\bf s}^*} \lambda_{\bf s}\,.
\label{afour}
\eea
We now have Eqs.~(\ref{first}), (\ref{aone}), and (\ref{afour}) in
four variables $\sum_{{\bf s} \neq {\bf 0},{\bf s}^*} \kappa_{\bf s}^2
\lambda_{\bf s}^2$, $\sum_{{\bf s} \neq {\bf 0},{\bf s}^*}
\lambda_{\bf s}$, $\lambda$, and $q_0-q_1$.
We can perform standard eliminations and obtain an expression for the
minimum of $q_0-q_1$:
\beq
	q_0 - q_1 = 2 {\sqrt{(1+C)q_0^2+C} - q_0 \over C}\,,
\label{simpler}
\eeq
where $C=\sum_{{\bf s} \neq {\bf 0},{\bf s}^*} 1/\kappa_{\bf s}^2$.

To reexpress Eq.~(\ref{simpler}) in the notation of the theorem
statement, we use $q_0 = 2 p_{0|0} - 1$ and $q_1= 2 p_{1|1}-1$, which
follow from Eqs.~(\ref{def2}) and (\ref{def3}).  We have
\beq
	p_{0|0} + p_{1|1} - 1 \geq 
	{\sqrt{(1+C)( 2 p_{0|0} - 1)^2 + C } - (2 p_{0|0} - 1) \over C}.
\eeq
This can be simplified by changing variables $y=p_{0|0}+p_{1|1}$ and
$x=p_{0|0}-p_{1|1}$ and solving for $y$.  We obtain
\beq
	p_{0|0} + p_{1|1} - 1 \geq 
	{\sqrt{1+(p_{0|0}-p_{1|1})^2} \over \sqrt{C+1} }\,.
\label{final}
\eeq
To achieve the desired lowest minimum in Eq.~(\ref{minmax}), we will
replace $C$ by its upper bound.  Since $\kappa_{\bf s} \geq 1$,
\beq
C\leq 16^n -2\,,\label{lcbound}
\eeq
and we obtain the statement Eq.~(\ref{curve}) to be proven.

The analysis for the cases when $a_{0{\bf s}}=a_{1{\bf s}}$ for some
{\bf s} is similar to the one just presented.  We obtain values of
$q_0-q_1$ which are always greater than in Eqs. (\ref{simpler}) and
(\ref{lcbound}), so these cases can be excluded.
\end{proof}

We remark that the weaker result, Eq.~(\ref{corr}), can be proved 
without the Lagrange multiplier analysis.  
When we maximize over all possible {\bf s}, the best achievable
$p_{0|0} + p_{1|1} - 1$ is at least, using Eq.~(\ref{minmax}):
\bea
        \max_{\bf s} {1 \over 2} |a_{0{\bf s}} - a_{1{\bf s}}|
        & \geq & 
        \max_{{\bf s}:a_{0{\bf s}} a_{1{\bf s}} < 0} 
        {1 \over 2} |a_{0{\bf s}} - a_{1{\bf s}}|
\nonumber
\\
\nonumber
\\      & \geq & 
        \max_{{\bf s}:a_{0{\bf s}} a_{1{\bf s}} < 0} 
        \sqrt{|a_{0{\bf s}} \, a_{1{\bf s}}|}
\nonumber
\\
        & = & \sqrt{\left| \rule{0pt}{2.2ex} \right. 
        \min_{{\bf s}:a_{0{\bf s}} a_{1{\bf s}} < 0}
        a_{0{\bf s}} \, a_{1{\bf s}} \left. \rule{0pt}{2.2ex} \right|}
\nonumber
\\
        & \geq & {1 \over \sqrt{16^n - 1}}\,.
\label{shortpf}
\eea
In the above proof, we use the orthogonality condition $\sum_{{\bf s}
\neq 0} a_{0{\bf s}} \, a_{1{\bf s}} = -1$, so that 
$\{{\bf s}:a_{0{\bf s}} \, a_{1{\bf s}} < 0\}$ is non-empty and 
$\min_{{\bf s} \neq 0} a_{0{\bf s}} \, a_{1{\bf s}} \leq {-1 \over 16^n-1}$.
Equation~(\ref{shortpf}) is independent of $\rho_{0,1}$, thus no further 
minimization over $\rho_0$ and $\rho_1$ is needed.

\section{An optimal LOCC protocol to distinguish $\tau_0$ and $\tau_1$}

\label{sec:talapp}

In this appendix, we describe and discuss an LOCC protocol to
distinguish $\tau_0$ from $\tau_1$ that achieves the bound in
Eq.~(\ref{talbdd}).
We employ the {\em Bloch representation} of a qubit.  
We identify the qubit state 
$\cos \theta \, |0\> + \sin \theta \, |1\>$ with the density matrix  
${1 \over 2} [I + \sin(2 \theta) \, \sigma_x + \cos (2 \theta) \, \sigma_z]$.  
The coefficients of $\sigma_x$ and $\sigma_z$ can be conveniently plotted 
as a 2-dimensional vector.
In this representation, orthogonal vectors are {\em anti-parallel}. 
The protocol is as follows:
\medskip
\begin{enumerate}
\item 
Alice first projects her qubit onto one of the two states $\eta_{\pm}
= {1 \over 2}[I \pm {1 \over \sqrt{2}} (- \sigma_x + \sigma_z)]$, and 
sends the result to Bob.  
\item 
Let $\cos \alpha = {1 \over \sqrt{3}} (1 + {1 \over \sqrt{2}})$ 
and $\sin \alpha = {1 \over \sqrt{3}} (1 - {1 \over \sqrt{2}})$. \\ 
If Alice obtains $\eta_+$, Bob projects his qubit onto the states 
$\eta_{+\pm} = {1 \over 2}[I \pm ( \cos \alpha \, \sigma_x + 
\sin \alpha \, \sigma_z)]$. \\
If Alice obtains $\eta_-$, Bob projects his qubit onto the states 
$\eta_{-\pm} = {1 \over 2}[I \pm ( \sin \alpha \, \sigma_x + 
\cos \alpha \, \sigma_z)]$.
\end{enumerate}
\medskip
It is straightforward to verify that this protocol achieves the bound
given by Eq.~(\ref{talbdd}).

\begin{figure}[http]
\centerline{\mbox{\psfig{file=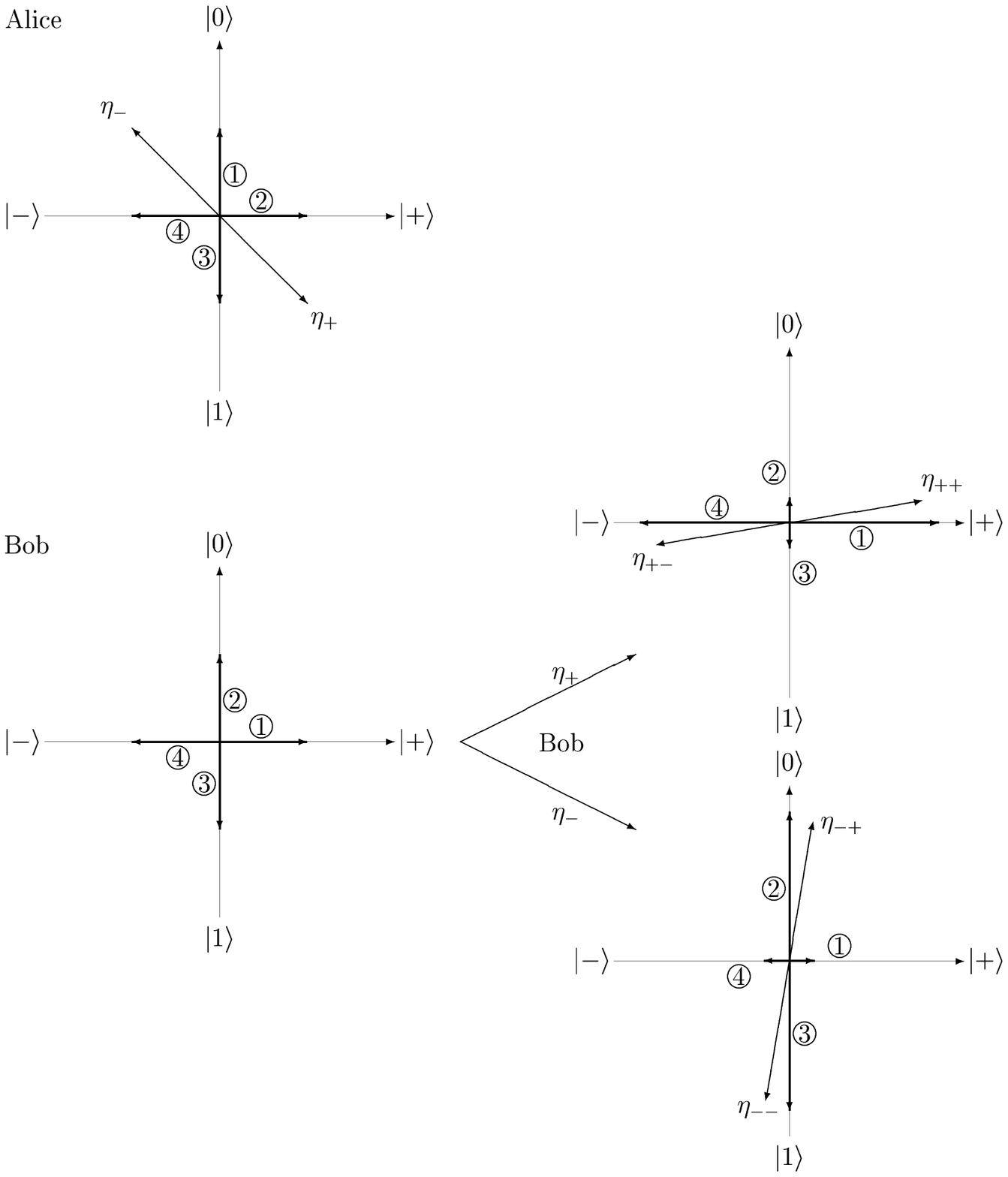, width=3.4in}}}
\caption{A protocol to distinguish $|0+\>, |$$+$$0$$\>, |11\>, |-$$-\>$, 
labeled by 
$\bigcirc \hspace*{-1.8ex} 1 \,$, 
$\bigcirc \hspace*{-1.8ex} 2 \,$,
$\bigcirc \hspace*{-1.8ex} 3 \,$,
$\bigcirc \hspace*{-1.8ex} 4 \,$.  
The classically correlated initial states of Alice and Bob are plotted
in the two left diagrams.  We will illustrate the case of equal prior
probabilities, indicated by the equal lengths of the four state
vectors.  Alice projects her qubit onto $\eta_\pm$, which
distinguishes $|0+\>$, $|$$-$$-$$\>$ ($\bigcirc \hspace*{-1.8ex} 1 \,$
and $\bigcirc \hspace*{-1.8ex} 4 \,$) from $|$$+$$0$$\>$, $|11\>$
($\bigcirc \hspace*{-1.8ex} 2 \,$ and $\bigcirc \hspace*{-1.8ex} 3
\,$) with high probability.  The two right diagrams represent the
qubit state of Bob conditioned on the two measurement outcomes of
Alice.  The conditional probabilities of the states are represented by
their lengths.
The optimal measurements of Bob to distinguish $\tau_0$ from
$\tau_1$ are given by the projections along $\eta_{\pm\pm}$.  
In fact, conditioned on $\eta_+$, 
$\eta_{++}$ is the projector along the direction of vector
sum of $\bigcirc
\hspace*{-1.8ex} 1 \,$ and $\bigcirc \hspace*{-1.8ex} 2 \,$, and 
similarly $\eta_{++}$ is the projector along the direction of vector
sum of $\bigcirc
\hspace*{-1.8ex} 3 \,$ and $\bigcirc \hspace*{-1.8ex} 4 \,$, which
explains the optimality.  Similar reasoning applies for $\eta_{-\pm}$.
Bob infers $\tau_0$ from measuring $\eta_{\pm +}$, and $\tau_1$ from
measuring $\eta_{\pm -}$.}
\label{taldec}
\end{figure}

The intuition behind this protocol is as follows.
It actually distinguishes among the four states $|$$0$$+$$\>$,
$|$$+$$0$$\>$, $|$$1$$1$$\>$, $|$$-$$-$$\>$ with high probability.
The first measurement extracts {\em no} information on whether the state is
$\tau_{0}$ or $\tau_{1}$.
Rather, it distinguishes $\{|0+\>$, $|$$-$$-$$\>\}$ from $\{|$$+$$0$$\>$,
$|11\>\}$ with high probability.  Then, Bob adaptively 
measure {\em approximately} along the $\{|+\>,|-\>\}$ or the
$\{|0\>,|1\>\}$ bases.
Bob's optimal measurement bases, the states $\eta_{\pm\pm}$, are
slightly tilted from $|+\>,|-\>,|0\>,|1\>$ to account for the 
imperfection of the inference from Alice's outcome.
A detailed pictorial explanation is given in Fig.~\ref{taldec}.

The POVM elements of this measurement are given by 
$M_0 = \eta_+ \otimes \eta_{++} + \eta_- \otimes \eta_{-+}$ and 
$M_1 = \eta_+ \otimes \eta_{+-} + \eta_- \otimes \eta_{--}$.  
They are not symmetric under the operation discussed in
Section~\ref{sec:tal}, ${\cal T}[\rho] = {1 \over 4} (\rho + {\sc s} \rho
{\sc s} + {\sc h}_2 \rho {\sc h}_2 + {\sc s} {\sc h}_2 \rho {\sc h}_2
{\sc s})$.
For this particular case, the POVM defined by ${\cal T}[M_0]$, ${\cal
T}[M_1]$ is also LOCC -- Alice and Bob flip two fair coins, which
determine if they are either to carry out the original protocol, or to
exchange their roles, or to carry out the protocol in the conjugate
basis, or to do both.
Note that ${\cal T}$ is {\em not} an LOCC operation, yet it always
transforms one LOCC POVM to another.
This example also illustrates that such symmetrizing operations on the
POVM elements, though they originate from symmetries of the states, do
not correspond to actual operations on the state.
We do not know if the class of LOCC POVMs is preserved under all
completely positive operations $\cal T$ such that ${\cal
T}^\dagger[\rho]$ is PPT if $\rho$ is PPT; in fact, we do not even
know if this class is preserved under LOCC operations.

\begin{figure}[f]
\begin{center}
\centerline{\mbox{\psfig{file=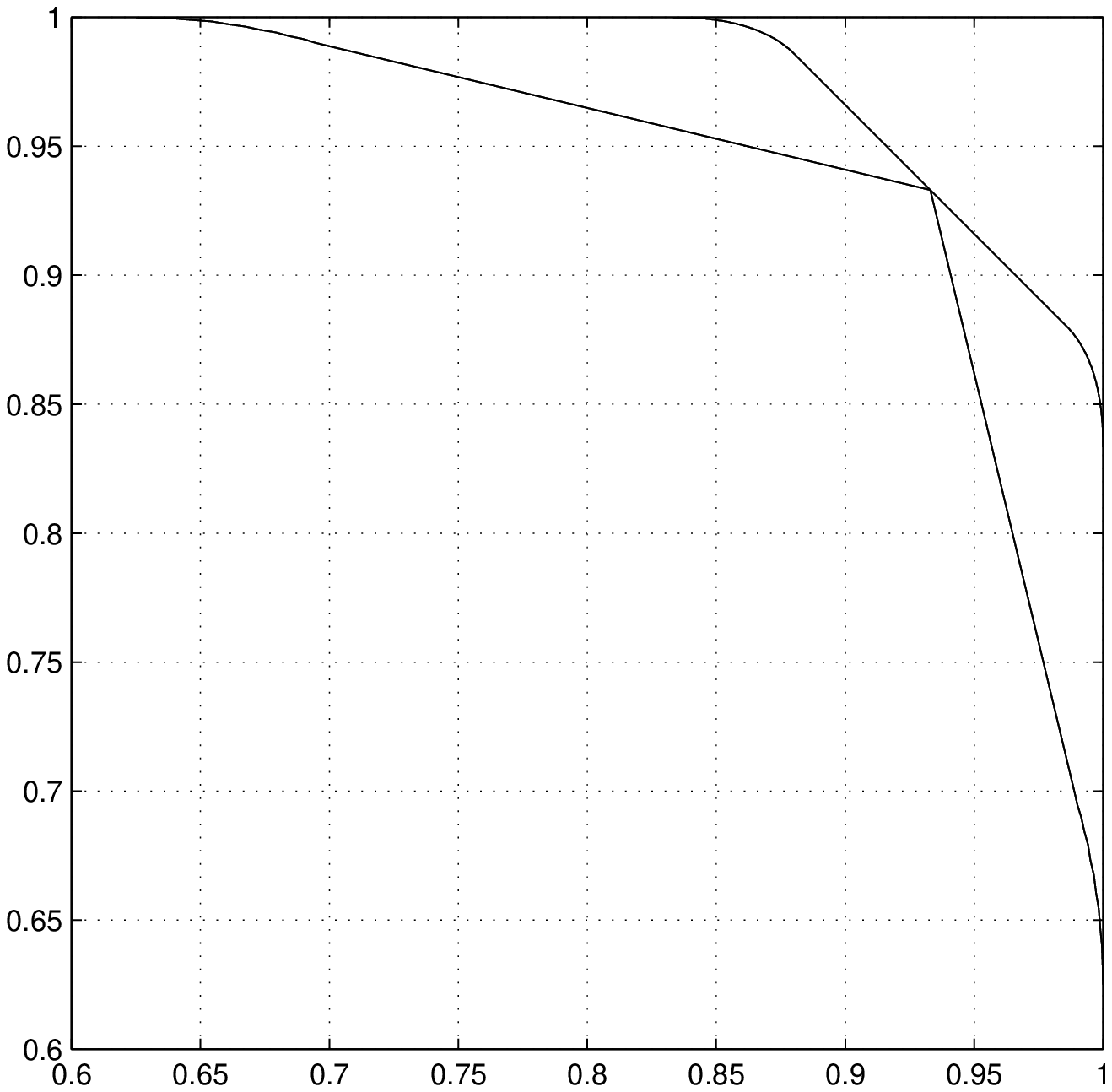,width=3in}}}
\vspace*{0.4cm}
\caption{The inner curve bounds the region of  
$(p_{0|0}^{(1)}, p_{1|1}^{(1)})$ attained by LOCC measurement
on states $\tau_{0,1}$, and the
outer curve bounds the region attained by PPT-preserving
measurements (from Fig. \protect\ref{fig:talppt}).}
\label{tallocc}
\end{center}
\end{figure}

Concerning the general question of whether PPT preserving POVMs are
achievable by LOCC, very little is presently known.  However, we have
been able to prove that, in ${\cal H}_2 \otimes {\cal H}_2$, PPT
preserving POVM measurements with two orthogonal POVM elements are
always in the LOCC class.  Applying this result to the $\tau_{0,1}$
measurement, we obtain values of $(p_{0|0}^{(1)},p_{1|1}^{(1)})$ that
we know to be achievable by LOCC, plotted in Fig.~\ref{tallocc}.

\bibliographystyle{IEEEbib}
\bibliography{refs}

\end{document}